\documentclass[preprint]{aastex}
\newcommand{\HI}{{\sc H}\,{\scriptsize{\sc I}}}
\usepackage{natbib}
\usepackage{ulem}
\usepackage{color}

\begin{document}
\title{Fermi LAT study of cosmic-rays and the interstellar medium in nearby
molecular clouds
}
\author{
M.~Ackermann\altaffilmark{1}, 
M.~Ajello\altaffilmark{2}, 
A.~Allafort\altaffilmark{2}, 
L.~Baldini\altaffilmark{3}, 
J.~Ballet\altaffilmark{4}, 
G.~Barbiellini\altaffilmark{5,6}, 
D.~Bastieri\altaffilmark{7,8}, 
K.~Bechtol\altaffilmark{2}, 
R.~Bellazzini\altaffilmark{3}, 
B.~Berenji\altaffilmark{2}
% ...
}
\altaffiltext{1}{Deutsches Elektronen Synchrotron DESY, D-15738 Zeuthen, Germany}
\altaffiltext{2}{W. W. Hansen Experimental Physics Laboratory, Kavli Institute for Particle Astrophysics and Cosmology, Department of Physics and SLAC National Accelerator Laboratory, Stanford University, Stanford, CA 94305, USA}
\altaffiltext{3}{Istituto Nazionale di Fisica Nucleare, Sezione di Pisa, I-56127 Pisa, Italy}
\altaffiltext{4}{Laboratoire AIM, CEA-IRFU/CNRS/Universit\'e Paris Diderot, Service d'Astrophysique, CEA Saclay, 91191 Gif sur Yvette, France}
\altaffiltext{5}{Istituto Nazionale di Fisica Nucleare, Sezione di Trieste, I-34127 Trieste, Italy}
\altaffiltext{6}{Dipartimento di Fisica, Universit\`a di Trieste, I-34127 Trieste, Italy}
\altaffiltext{7}{Istituto Nazionale di Fisica Nucleare, Sezione di Padova, I-35131 Padova, Italy}
\altaffiltext{8}{Dipartimento di Fisica ``G. Galilei", Universit\`a di Padova, I-35131 Padova, Italy}
% ...

\begin{abstract}

We report an analysis of the interstellar $\gamma$-ray emission from 
the Chamaeleon, R~Coronae Australis (R CrA), and Cepheus and Polaris flare 
regions with the {\it Fermi} Large Area Telescope. They are
among the nearest molecular cloud complexes, within $\sim$ 300 pc from the 
solar system. The $\gamma$-ray emission produced by interactions of 
cosmic-rays (CRs) and interstellar gas in those molecular clouds is useful 
to study the CR densities and distributions of molecular gas
close to the solar system. The obtained $\gamma$-ray emissivities above 250 MeV are 
(5.9 $\pm$ 0.1$_{\rm stat}$ $^{+0.9}_{-1.0}$$_{\rm sys}$) 
$\times$ 10$^{-27}$ photons s$^{-1}$ sr$^{-1}$ H-atom$^{-1}$, 
(10.2 $\pm$ 0.4$_{\rm stat}$ $^{+1.2}_{-1.7}$$_{\rm sys}$) 
$\times$ 10$^{-27}$ photons s$^{-1}$ sr$^{-1}$ H-atom$^{-1}$,
and (9.1 $\pm$ 0.3$_{\rm stat}$ $^{+1.5}_{-0.6}$$_{\rm sys}$) 
$\times$ 10$^{-27}$ photons s$^{-1}$ sr$^{-1}$ H-atom$^{-1}$ 
for the Chamaeleon, R~CrA, and Cepheus and Polaris flare regions, respectively. 
Whereas the energy dependences of the emissivities agree well with that 
predicted from direct CR observations at the Earth,
the measured emissivities from 250 MeV to 10 GeV indicate a variation of the CR
density by $\sim$ 20 \% in the neighborhood of the solar system,
even if we consider systematic uncertainties.
The molecular mass calibrating ratio, 
$X_{\rm CO} = N({\rm H_{2}})/W_{\rm CO}$, is found to be 
(0.96 $\pm$ 0.06$_{\rm stat}$ $^{+0.15}_{-0.12}$$_{\rm sys}$) $\times$10$^{20}$
H$_2$-molecule cm$^{-2}$ (K km s$^{-1}$)$^{-1}$, 
(0.99 $\pm$ 0.08$_{\rm stat}$ $^{+0.18}_{-0.10}$$_{\rm sys}$) $\times$10$^{20}$
H$_2$-molecule cm$^{-2}$ (K km s$^{-1}$)$^{-1}$, and
(0.63 $\pm$ 0.02$_{\rm stat}$ $^{+0.09}_{-0.07}$$_{\rm sys}$) $\times$10$^{20}$
H$_2$-molecule cm$^{-2}$ (K km s$^{-1}$)$^{-1}$
for the Chamaeleon, R~CrA, and Cepheus and Polaris flare regions, respectively,
suggesting a variation of $X_{\rm CO}$ in the vicinity of the solar system.
From the obtained values of $X_{\rm CO}$, the masses of molecular gas traced 
by $W_{\rm CO}$ in the Chamaeleon, R~CrA, and Cepheus and Polaris flare regions are 
estimated to be $\sim$ 5$\times$10$^{3}$ $M_{\odot}$, $\sim$ 10$^{3}$ $M_{\odot}$, 
and $\sim$ 3.3$\times$$10^{4}$ $M_{\odot}$, respectively. A comparable amount of 
gas not traced well by standard \HI\ and CO surveys is found in the regions investigated.

\end{abstract}


\section{Introduction}

Observations of high-energy $\gamma$-ray emission ({\it E} $\gtrsim$ 30 MeV) from
molecular clouds can be used to study the cosmic-ray (CR) production,
the CR density, and the distribution of the interstellar medium (ISM) in such
systems. $\gamma$-rays are produced in the ISM by interactions of
high-energy CR protons and electrons with the interstellar gas, via 
nucleon-nucleon collisions, electron Bremsstrahlung, and inverse Compton
(IC) scattering. Since the $\gamma$-ray production cross section is
almost independent of the chemical or thermodynamic state of the ISM,
and the interstellar gas is essentially transparent to those high-energy 
photons, observations in $\gamma$-rays have been recognized as a
powerful probe of the distribution of interstellar matter. If the gas
column densities are estimated with good accuracy by observations in
other wavebands such as radio, infrared, and optical, the CR spectrum and 
density distributions can be examined as well. Molecular clouds that are 
within 1 kpc from the solar system (namely nearby molecular clouds) and 
have masses greater than a few 10$^3$ {\it M}$_{\odot}$ are well suited 
for an analysis of their $\gamma$-ray emission to 
investigate the distribution of CR densities and interstellar gas since
they are observed at high latitudes and therefore largely free 
from confusion with the strong emission from the Galactic plane.  
Study of such nearby molecular clouds in $\gamma$-rays can be dated back to
the COS-B era (e.g., Bloemen et al. 1984) and was advanced by the EGRET 
on board {\it Compton Gamma-Ray Observatory} (e.g., Hunter et
al. 1994). Although some important information has been obtained on
properties of CRs and the ISM by these early observations, detailed
studies have only been performed on giant molecular clouds with masses 
greater than $\sim$ 10$^5$ {\it M}$_{\odot}$ such as the Orion complex 
(e.g., Digel et al. 1999). The data above 1 GeV, which are crucial to
study CR nuclei spectra, suffered from the limited photon statistics, 
angular resolution, and energy coverage of these early missions.

The advent of the {\it Fermi} Gamma-ray Space Telescope launched in 2008
has improved the situation significantly. The sensitivity of the LAT 
(Large Area Telescope) on board {\it Fermi} 
is more than an order of magnitude better than that of the EGRET, and enables 
resolving more point sources and studying the diffuse $\gamma$-ray
emission with unprecedented sensitivity. In addition, newer surveys 
of the ISM (e.g., Dame et al. 2001, 
Kalberla et al. 2005, and Grenier et al. 2005) allow us to investigate the CR
spectral and density distributions with better accuracy.

Here we report a {\it Fermi} LAT study of diffuse $\gamma$-rays from
the Chamaeleon, R Coronae Australis (R CrA), and Cepheus and Polaris flare molecular clouds. 
They are among the nearest ($\lesssim$ 300 pc from the solar system)
molecular clouds exhibiting star formation activity. Although 
EGRET observed $\gamma$-ray emission associated with the molecular gas 
in the Chamaeleon region (Grenier et al. 2005), no detailed study of 
CR and matter distributions for the Chamaeleon and R CrA regions has been performed yet 
since they have rather small masses 
($\lesssim$ 10$^{4}$ {\it M}$_{\odot}$, about 1/10 of that of the Orion 
molecular cloud) and consequently small $\gamma$-ray fluxes. 
We also analyzed in detail the region of the Cepheus and Polaris flares which
was included in the {\it Fermi} LAT study of the second Galactic quadrant (Abdo et al. 2010b).
It is located in the direction almost opposite to the Chamaeleon region in the 
Gould Belt (see, e.g., Perrot and Grenier 2003), therefore we can investigate
the distribution of the CR density over several hundred pc but still inside
the coherent environment of the Gould Belt. 
This paper is organized as follows. We first describe the observations as well as
the sky model preparation and the data analysis in Section
\ref{sec:Data_analysis}, and show the obtained results in Section 
\ref{sec:Results}. We then discuss the CR and matter distributions in 
Section \ref{sec:Discussion} and give conclusions in Section 
\ref{sec:Summary_and_conclusions}.

\section{Data Analysis}
\label{sec:Data_analysis}

\subsection{Observations and Data Reduction}
\label{sec:ObservationAndDataReduction}

The LAT, on board the {\it Fermi} Gamma-ray Space Telescope, is a
pair-tracking detector to study $\gamma$-rays from $\sim$ 20 MeV
to more than 300 GeV. It consists of an array of 4$\times$4 
conversion and tracking modules built with 
tungsten foils and silicon microstrip detectors to measure the arrival 
directions of incoming $\gamma$-rays and a hodoscopic cesium iodide 
calorimeter to determine the photon energies. The modules are surrounded 
by 89 segmented plastic scintillators serving as an anticoincidence
detector to reject charged-particle background events. A detailed description of the 
LAT instrumentation can be found in Atwood et al. (2009) and the 
on-orbit calibration is discussed in Abdo et al. (2009a). 

Science operations with the LAT started on 2008 August 4. For this
analysis we have accumulated events obtained from 2008 August 4 to 2010
May 9. During this time interval the LAT was operated in sky survey mode 
nearly all of the time and scanned the $\gamma$-ray sky with relatively 
uniform exposure over time (within 10\% in regions studied). 
We used the standard LAT analysis software, 
Science Tools\footnote{Available from the {\it Fermi} Science Support Center
(http://fermi.gsfc.nasa.gov/ssc/).} version v9r16p0 and the response 
function P6\_V3\_DIFFUSE, which was developed to account for the
detection inefficiencies due to pile-up and accidental coincidence of 
events (Rando et al. 2009). We applied the following event selection 
criteria to the $\gamma$-ray events: (1) events must satisfy 
the standard low-background event selection (so-called diffuse class events;
Atwood et al. 2009), (2) the reconstructed zenith angles of the arrival
direction of photons are less than 100$^{\circ}$ in order to reduce 
contamination of photons from the bright Earth rim, and (3) the center
of the LAT field of view is within 52$^{\circ}$ from the zenith
direction of the sky, in order to exclude the data obtained during the 
relatively short time intervals of pointed observations when the rocking 
angle of the LAT was larger. The exposure maps were generated with the 
same 100$^{\circ}$ limit on zenith angle for each direction in the sky.

\begin{figure}
\begin{minipage}{0.5\hsize}
\includegraphics[width=80mm]{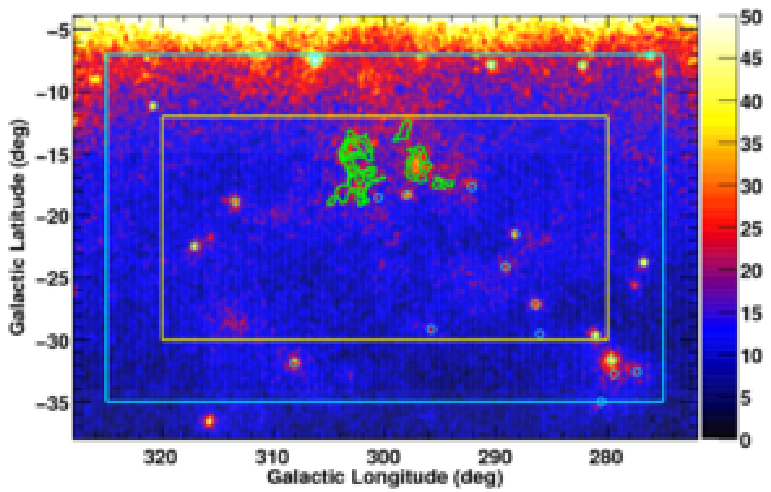}
\end{minipage}
\begin{minipage}{0.5\hsize}
\includegraphics[width=80mm]{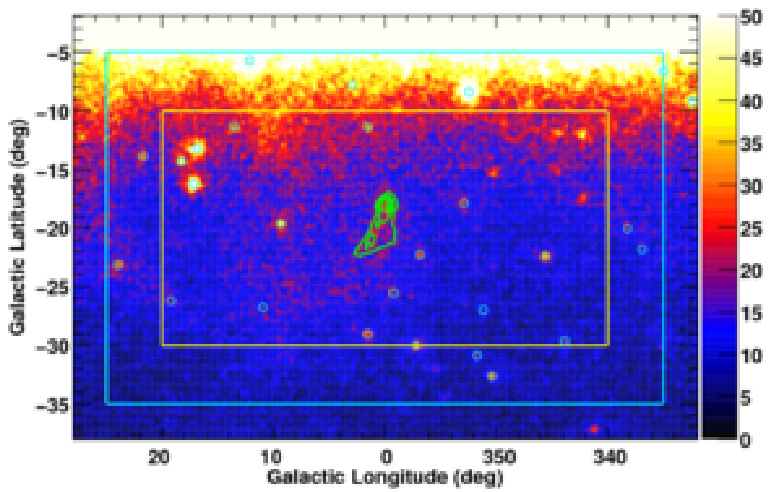}
\end{minipage}\\
\begin{minipage}{0.5\hsize}
\includegraphics[width=80mm]{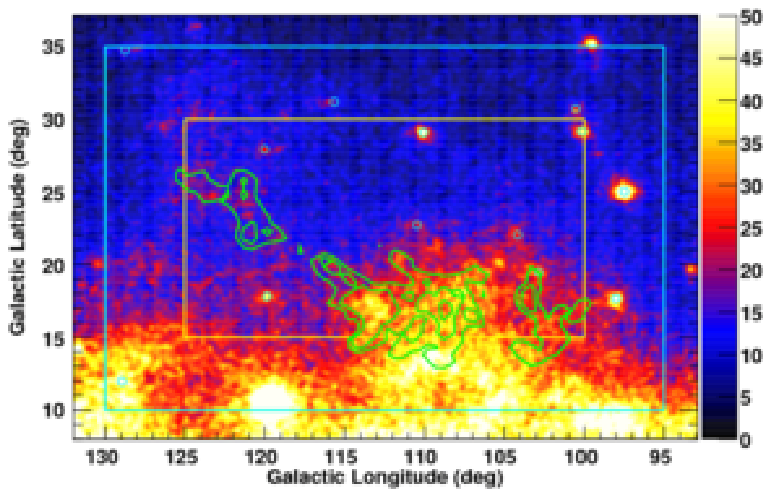}
\end{minipage}
\caption{$\gamma$-ray count maps above 250 MeV for the Chamaeleon (top left), 
R CrA (top right), and Cepheus and Polaris flare (bottom left) regions, smoothed 
with a Gaussian of a standard deviation $\sigma =$ 0.5$^{\circ}$ for display. The contours indicate
intensities $W_{\rm CO}$ of the 2.6 mm line of CO (with the levels of 4, 8,
12, and 16 K km s$^{-1}$) by Dame et al (2001), as a standard tracer of the molecular 
gas. Cyan circles show the positions of point sources with high significance 
(TS $\geq$ 50) in the First {\it Fermi} LAT catalog (1FGL) by Abdo et al. (2010a). 
The yellow squares indicate the ROI analyzed for each of the regions. Point sources 
outside of this ROI but inside the cyan square are taken into account in the analysis.}
\label{fig:Cham_R_CrA_CePo_image}
\end{figure}

The count maps obtained ($E > 250 $ MeV) in the Chamaeleon, R CrA, and Cepheus
and Polaris flare regions are shown in Figure \ref{fig:Cham_R_CrA_CePo_image}: 
we set the lower energy limit at 250 MeV 
to utilize good angular resolution (68\% containment 
radius is $\lesssim$ 1.5$^{\circ}$ above 250 MeV)
and the upper energy limit at 10 GeV because of limited photon statistics.
We also show positions of point sources with high significance (test
statistic, TS\footnote{TS is defined as 
${\rm TS} = 2({\rm ln}L-{\rm ln}L_{0})$, where $L$ and $L_{0}$ are the
maximum likelihoods obtained with and without the source included in the model
fitting, respectively; see Mattox et al. (1996).}, 
greater than 50) and 2.6 mm carbon-monoxide CO line intensities on the
maps. The yellow square indicates the region of interest (ROI) analyzed
for each of the regions. In order to take into account the spillover
from point sources outside of the ROIs, we also included point sources
lying just outside ($\leq$ 5$^{\circ}$) of the region boundaries. 
Contamination due to the diffuse emission from the interstellar gas
outside of the ROIs is also taken into account through the convolution 
of maps larger than the ROIs. 

\subsection{Model Preparation}
\label{sec:Model_preparation}

Since the ISM is optically thin to $\gamma$-rays in the energy range 
considered in the paper and the CR spectrum is
not expected to vary significantly within small regions, the
$\gamma$-ray intensity from CR protons and electrons interacting with
the interstellar gas may be modeled as a sum of emission from separate
gas phases (e.g., Lebrun et al. 1983). This approach has been successfully
applied in recent studies of diffuse $\gamma$-rays by the LAT (e.g.,
Abdo et al. 2010b and Ackermann et al. 2011). We followed this method and
prepared template maps as described below.

\subsubsection{H {\scriptsize{I}} and CO Maps}
\label{sec:HI_map}

\begin{figure}
      \begin{minipage}{0.5\hsize}
      \includegraphics[width=80mm]{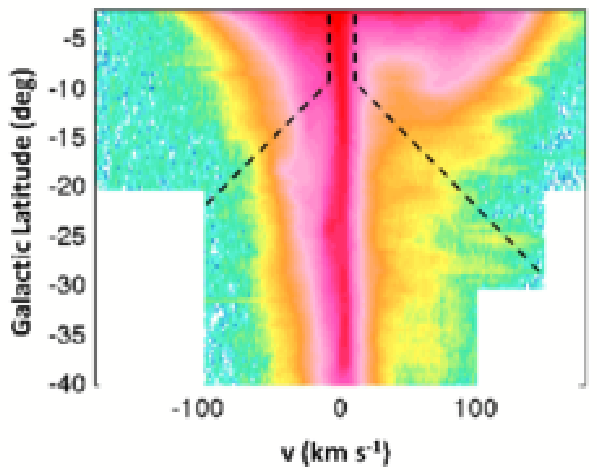}
      \end{minipage}
      \begin{minipage}{0.5\hsize}
      \includegraphics[width=80mm]{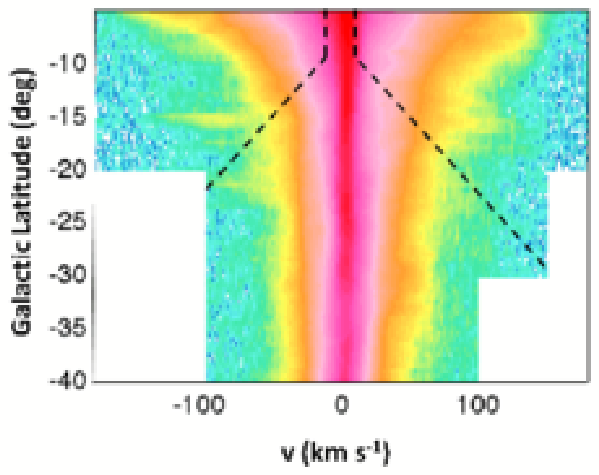}    
      \end{minipage}\\
      \begin{minipage}{0.5\hsize}
      \includegraphics[width=80mm]{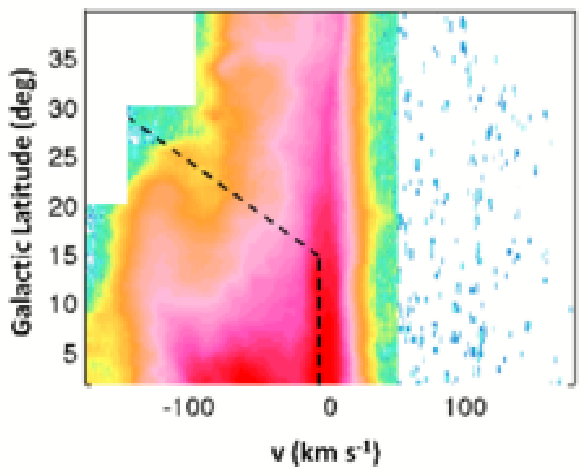}
      \end{minipage}
      \begin{minipage}{0.5\hsize}
      \end{minipage} 
      \caption{Latitude-velocity diagrams of the intensity of the 21 cm line 
      (log scale in atom cm$^{-2}$) for the Chamaeleon (top left), R CrA 
      (top right), and Cepheus and Polaris flare (bottom left) regions. The dashed
      lines indicate the region boundary between the local 
      region ($\lesssim$ 300 pc) and the rest.}
      \label{fig:bv_Cham_R_CrA_CePo}
      \end{figure}

      We calculated the column densities {\it N}({\HI}) of atomic hydrogen from the 
      Leiden/Argentine/Bonn Galactic {\HI} survey by Kalberla et al. (2005). 
      The optical depth correction of the {\HI} gas is applied under 
      the assumption of a uniform spin temperature {\it T}${\rm _S}$ = 125 K, 
      the value which has often been used in previous studies 
      (e.g., Abdo et al. 2009c and Abdo et al. 2010b). 
      This choice of {\it T}${\rm _S}$ allows us to directly compare our 
      results with other studies. In
      order to evaluate systematic uncertainties due to the optical
      depth correction, we also tried several different choices of 
      {\it T}${\rm _S}$ as described in Section~\ref{sec:Results}. 
      We note that the true {\it T}${\rm _S}$ is likely not to be
      uniform even in small regions like the ones we are studying, but
      exploring a non-uniform {\it T}${\rm _S}$ is beyond the scope of this paper.
      We separated the {\HI} column densities in two regions along the line of sight; one
      corresponds to the local region ($\lesssim$ 300 pc) to take
      into account the ambient atomic gas surrounding the molecular
      cloud and the other corresponds to the rest to take into account
      the remaining gas along the line of sight.
      From the velocity distribution of the CO emission which traces 
      the molecular cloud, we determined 
      the boundary as shown in Figure \ref{fig:bv_Cham_R_CrA_CePo}.   
      The local velocity range is $-10 < v_{\rm LSR} < 15 $ km s$^{-1}$ for the Chamaeleon region 
      and  $-15 < v_{\rm LSR} < 15 $ km s$^{-1}$ for the R CrA region for 
      $b > -10^{\circ}$. Below $-10^{\circ}$, $|v_{\rm LSR}|$ 
      is increased to 80 km s$^{-1}$ at $b = -20.5^{\circ}$ in both regions,
      since the {\HI} gas at such high latitude is likely to be local.
      The obtained {\it N}({\HI}) maps are shown in the top panels
      of Figure \ref{fig:Cham_gas_model_map} and Figure 
      \ref{fig:RCrA_gas_model_map} for the Chamaeleon region and the R CrA region, respectively.
      For the Cepheus and Polaris flare region, the cut falls 
      in between the Gould Belt lines and the Local-Arm lines at 
      $-8$ km s$^{-1}$ at $b < 15^{\circ}$, and then opens up to $-100$ km s$^{-1}$ at $b = 24^{\circ}$. 
      Since the amount of {\HI} gas in the Local-Arm and beyond is 
      comparable to that in the Gould Belt for the Cepheus and Polaris flare 
      region in $15^{\circ} < b < 20^{\circ}$ (see also Figure 
      \ref{fig:CePo_gas_model_map}), using two {\HI} template maps is crucial.
      For the Chamaeleon and the R CrA regions, the non-local {\HI} gas is almost negligible.

      The integrated intensities of the 2.6 mm line, $W_{\rm CO}$,  
      have been derived from the composite survey of Dame, Hartmann, \& Thaddeus (2001). 
      We used this $W_{\rm CO}$ map as a standard    
      molecular-gas tracer. For better signal-to-noise ratio, the data
      have been filtered with the moment-masking technique 
      (Dame 2011) to reduce the noise while keeping the
      resolution of the original data. 
      Since most of the molecular gas turned out to be local  
      according to our velocity cuts, we used 
      only the local CO map in the $\gamma$-ray analysis.

\subsubsection{Excess $A{\rm v}$ Map}
\label{sec:Av_map}
   
      Dust is a commonly-used tracer of the neutral interstellar 
      gas. By comparing the $\gamma$-ray observations by EGRET with radio
      surveys and the dust thermal emission, Grenier et al. (2005)
      reported a considerable amount of gas at the interface between the
      atomic/molecular phases in the solar neighborhood, associated with
      cold dust but not properly traced by {\HI} and CO surveys.
      This finding was confirmed by dedicated analyses of the diffuse
      $\gamma$-ray emission with the {\it Fermi} LAT (Abdo et al. 2010b and
      Ackermann et al. 2011). In order to take into account this additional interstellar gas, 
      we constructed visual extinction ($A{\rm v}$) maps, based on the 
      extinction maps derived by Schlegel et al. (1998) from {\it IRAS} and COBE DIRBE data.
      The $A{\rm v}$ map on the assumption of a constant gas-to-dust ratio
      provides an estimate of the total column densities.  
      After fitting a linear combination of the 
      {\it N}({\HI}) and $W_{\rm CO}$ maps through a minimum
      sum-of-square-residuals criterion in each ROI separately, 
      we thus obtained a residual extinction map, 
      $A{\rm v}_{\rm res}$ map, accounting for the additional gas which
      is not properly traced by the {\HI} and CO surveys.
      Negative residuals are likely to be due to the fluctuation of the 
      original $A{\rm v}$ map. For simplicity, we clipped data around 0 in the $A{\rm v}_{\rm res}$ map.       

\begin{figure}[ht]
  \begin{minipage}{0.5\hsize}
    \includegraphics[width=80mm]{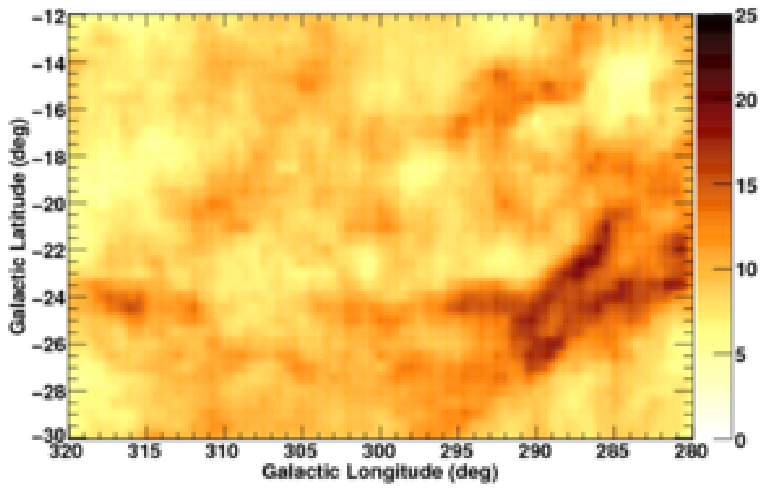}
    \label{fig:Cham_HIR12_map}
  \end{minipage}
  \begin{minipage}{0.5\hsize}
    \includegraphics[width=80mm]{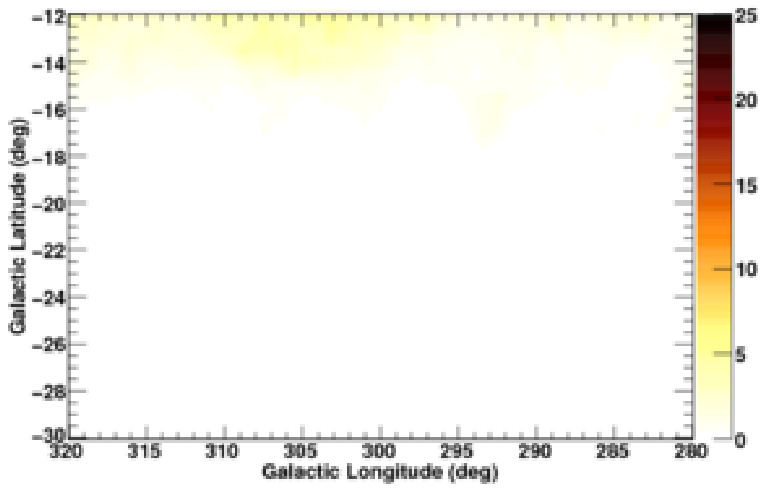}
    \label{fig:Cham_HIR0-R11_map}
  \end{minipage}\\
 \begin{minipage}{0.5\hsize}
    \includegraphics[width=80mm]{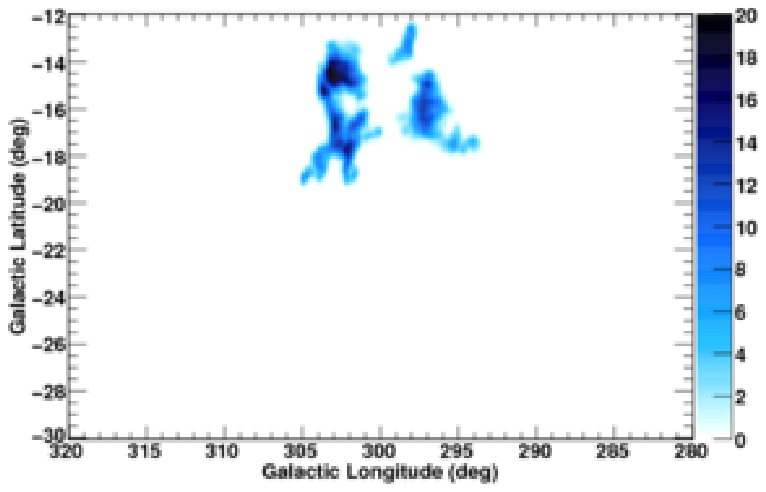}
    \label{fig:Cham_CO_map}
  \end{minipage} 
  \begin{minipage}{0.5\hsize}
    \includegraphics[width=80mm]{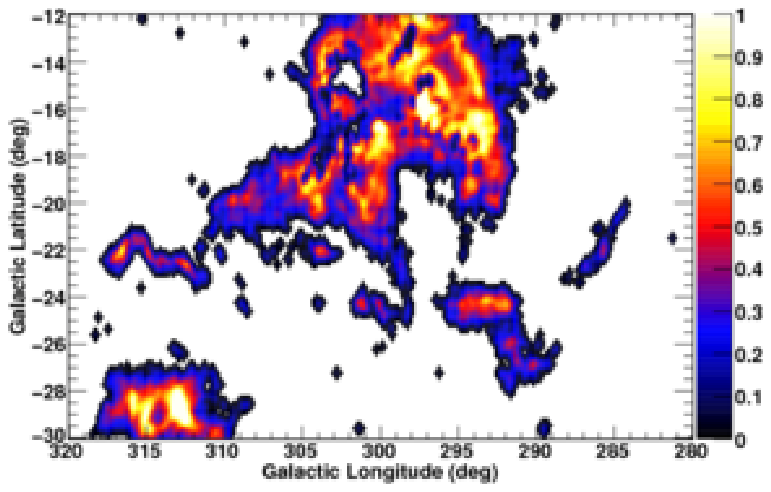}
    \label{fig:Cham_DG_map}
  \end{minipage}
 \caption{Template gas maps for the Chamaeleon region; {\it N}({\HI}) (local) 
 (top left) and {\it N}({\HI}) (non-local) (top right) in units of 
 10$^{20}$ H-atoms cm$^{-2}$, $W_{\rm CO}$ (bottom left) in units of 
 K km s$^{-1}$, and $A{\rm v}_{\rm res}$ (bottom right) in units of magnitudes. 
 The two {\it N}({\HI}) maps have been smoothed with a Gaussian of a standard deviation
 $\sigma = 1^{\circ}$ for display while the other two maps have been
 smoothed with a Gaussian of $\sigma = 0.25^{\circ}$ in order to keep fine
 structures seen in $W_{\rm CO}$ and $A{\rm v}_{\rm res}$ distributions.}
 \label{fig:Cham_gas_model_map}
\end{figure}

\begin{figure}
  \begin{minipage}{0.5\hsize}
    \includegraphics[width=80mm]{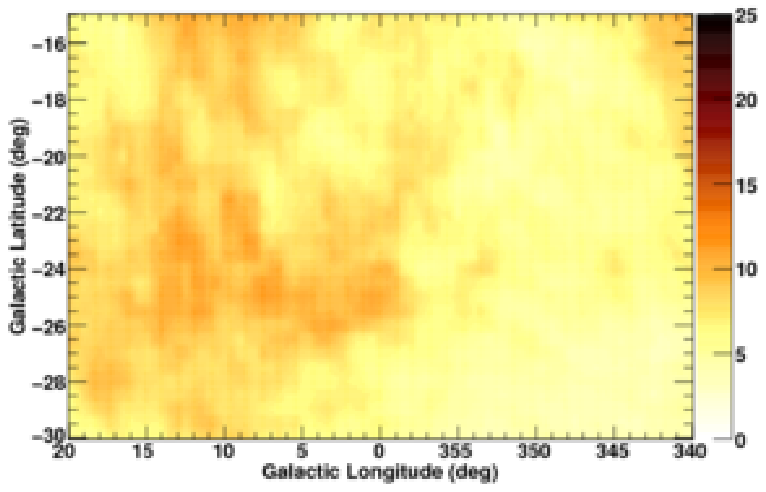}
   \label{fig:RCrA_HIR12_map}
   \end{minipage}
  \begin{minipage}{0.5\hsize}
  \end{minipage}\\
  \begin{minipage}{0.5\hsize}
    \includegraphics[width=80mm]{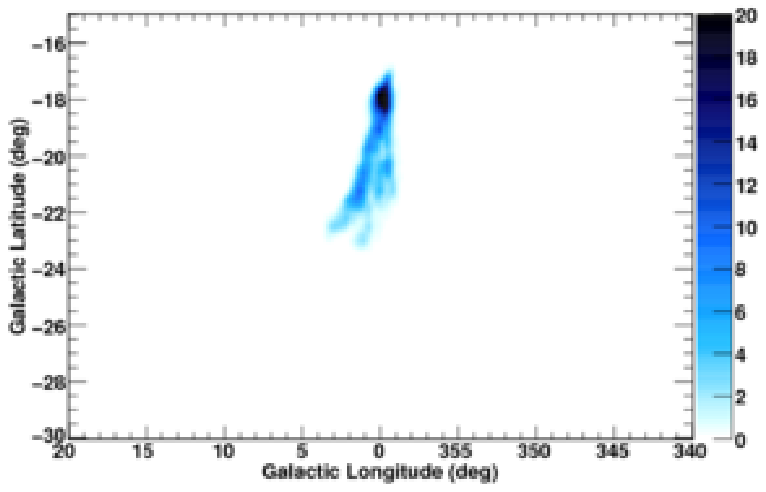}    
    \label{fig:RCrA_CO_map}
  \end{minipage}
  \begin{minipage}{0.5\hsize}
    \includegraphics[width=80mm]{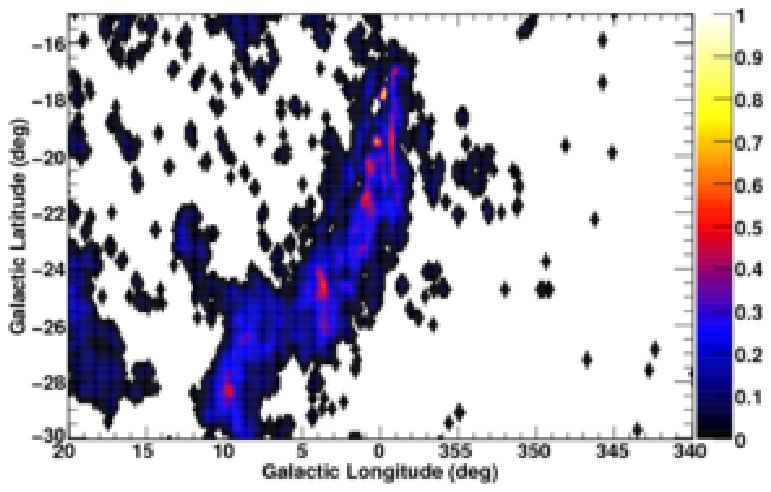}
   \label{fig:RCrA_DG_map}
  \end{minipage}
  \caption{Template gas maps for the R CrA region; 
 {\it N}({\HI}) (local) (top left) in units of 
 10$^{20}$ atoms cm$^{-2}$, $W_{\rm CO}$ (bottom left) in units of 
 K km s$^{-1}$, and $A{\rm v}_{\rm res}$ (bottom right) in units of
 magnitudes. They have been smoothed in the same way as maps in Figure 
 \ref{fig:Cham_gas_model_map}.  The non-local {\it N}({\HI}) is almost 
 0 in our ROI and hence is not shown here.} 
 \label{fig:RCrA_gas_model_map}
\end{figure}

\begin{figure}
  \begin{minipage}{0.5\hsize}
    \includegraphics[width=80mm]{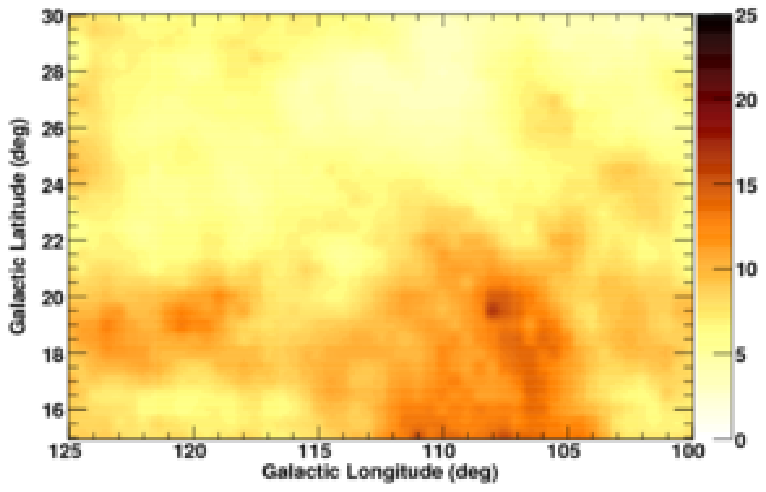}
   \label{fig:CePo_HIR12_map}
  \end{minipage}
  \begin{minipage}{0.5\hsize}
    \includegraphics[width=80mm]{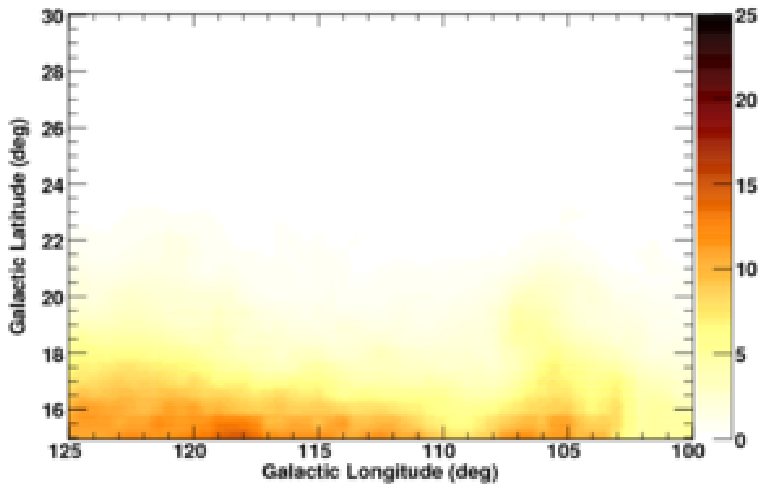}
    \label{fig:CePo_HIR0-R11_map}
  \end{minipage}\\
  \begin{minipage}{0.5\hsize}
    \includegraphics[width=80mm]{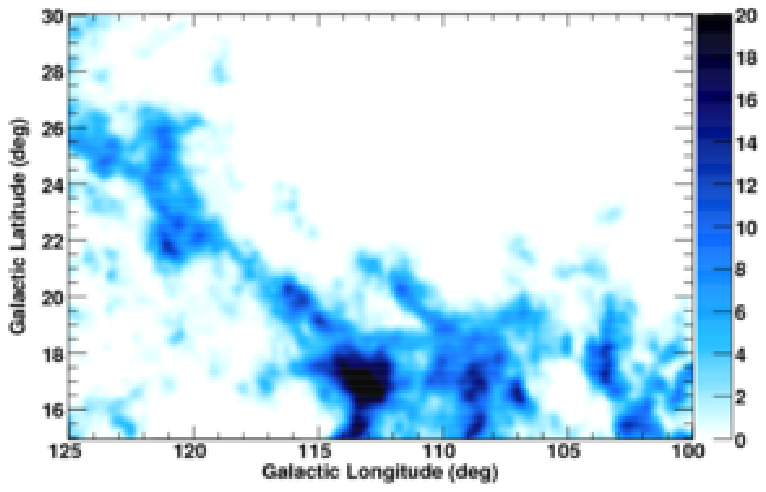}
   \label{fig:CePo_CO_map}
  \end{minipage} 
  \begin{minipage}{0.5\hsize}
    \includegraphics[width=80mm]{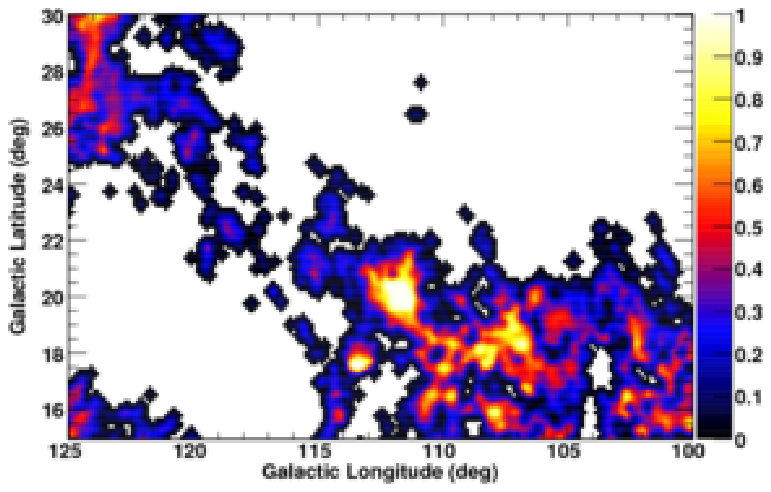}
   \label{fig:CePo_DG_map} 
  \end{minipage}
  \caption{Template gas maps for the Cepheus and Polaris flare region; 
 {\it N}({\HI}) (local) (top left) and {\it N}({\HI}) (non-local) (top right) 
 in units of 10$^{20}$ atoms cm$^{-2}$, $W_{\rm CO}$ (bottom left) in units of 
 K km s$^{-1}$, and $A{\rm v}_{\rm res}$ (bottom right) in units of
 magnitudes. They have been smoothed in the same way as maps in Figure 
 \ref{fig:Cham_gas_model_map}.} 
 \label{fig:CePo_gas_model_map}
\end{figure}

We present gas maps used for the analysis of the Chamaeleon, R CrA, and 
Cepheus and Polaris flare regions in Figures \ref{fig:Cham_gas_model_map}, 
\ref{fig:RCrA_gas_model_map}, and \ref{fig:CePo_gas_model_map}, respectively.

\clearpage

\subsubsection{IC, Point Sources and Isotropic Component}
\label{sec:IC_Iso_PS}

In addition to the gas-related components described above, we also need
to take into account $\gamma$-rays from IC scattering, contributions of 
point sources, extragalactic diffuse emission, and instrumental residual background. 
To model the IC emission, we used GALPROP 
(e.g., Strong \& Moskalenko 1998), a numerical code which solves the CR 
transport equation within our Galaxy and predicts the $\gamma$-ray
emission produced via interactions of CRs with the ISM. IC emission is
calculated from the distribution of propagated electrons and the
radiation field model developed by Porter et al. (2008). 
Here we adopt
the IC model map produced in the GALPROP run 54$\_$77Xvarh7S\footnote{http://www.mpe.mpg.de/\~{}aws/propagate.html} as a baseline 
model,
in which the CR electron spectrum is 
adjusted based on the {\it Fermi} LAT measurement.
In this model, the CR source distribution model is 
adjusted to the LAT data, and is somewhat more concentrated to the inner 
Galaxy than the pulsar distribution by Lorimer (2004).
To take into account uncertainties of the
CR electron spectrum and radiation field on the Galactic scale,
we set the normalization of this IC component free in each energy bin when we perform the fit
(see Section \ref{sec:Analysis_procedure}).
In order to evaluate the systematic uncertainty due to IC models,
we also tried four other models, which are constructed under
different assumptions about the distribution of 
CR sources such as supernova remnants (Case \& Bhattacharya 1998) and pulsars (Lorimer 2004), 
and intensity of the interstellar
radiation field depending on the input luminosity of the Galactic bulge component
(e.g., Ackermann et al. 2012).

To take into account $\gamma$-ray point sources, we used the 1FGL
catalog based on the first 11 months of the science phase of the mission 
(Abdo et al. 2010a). We included in our analysis point sources detected
with TS $\geq$ 50 in the 1FGL catalog and other significant point sources
included in the 2FGL catalog as described in Section \ref{sec:Analysis_procedure}. 

To represent the sum of the extragalactic diffuse emission and the residual 
background from the misclassified CR interactions in the LAT, we adopted
a publicly available isotropic spectrum\footnote{isotropic\_iem\_v02.txt 
from http://fermi.gsfc.nasa.gov/ssc/data/access/lat/BackgroundModels.html}
obtained by a fit to emission from the high latitude sky ($b >$ 30$^{\circ}$). 
This component is fixed in our analysis. The uncertainty due to this isotropic
term will be discussed in Section \ref{sec:Results}.

\clearpage

\subsection{Analysis Procedure}
\label{sec:Analysis_procedure}

With the usual assumptions of optical thinness and that CRs uniformly
thread the ISM, $\gamma$-ray intensity {\it I}$_\gamma (l,b)$ 
(s$^{-1}$ cm$^{-2}$ sr$^{-1}$ MeV$^{-1}$) at a given energy can be
modeled as 
\begin{eqnarray*}
I_{\gamma}(l,b) & = & \sum_{i=1}^{2} q_{{\rm H I},i} \cdot N({\rm
 H_{\ I}})(l,b)_{i} + q_{\rm CO} \cdot W_{\rm CO}(l,b) \\
                & + & q_{\rm Av} \cdot A{\rm v}_{\rm res}(l,b) +
	{\rm c_{IC}}\cdot I_{\rm IC}(l,b) + I_{\rm iso} + \sum_{j} {\rm PS}_{j}
                 \ \ \ \ \ \ \ \ \ \ \ \ \ \ \ \ (1)
\label{eq:fit_model} 
\end{eqnarray*}
where sum over {\it i} represents the two regions (local and non-local regions),  
{\it q}$_{{\rm {HI}},i}$ (s$^{-1}$ sr$^{-1}$ MeV$^{-1}$), {\it q}$_{{\rm {CO}}}$ 
(s$^{-1}$ cm$^{-2}$ sr$^{-1}$ MeV$^{-1}$ (K km s$^{-1}$)$^{-1}$), 
and {\it q}$_{\rm {Av}}$ (s$^{-1}$ cm$^{-2}$ sr$^{-1}$ MeV$^{-1}$ mag$^{-1}$) are the
emissivity per {\HI} atom, per $W_{\rm CO}$ unit, and per 
$A{\rm v}_{\rm res}$ magnitude, respectively. {\it I}$_{\rm IC}$ and {\it I}$_{\rm iso}$ 
are the IC model and isotropic background intensities 
(s$^{-1}$ cm$^{-2}$ sr$^{-1}$ MeV$^{-1}$),
respectively, and PS$_{j}$ represents contributions of point sources.
Hard $\gamma$-ray emission with a characteristic bubble shape 
above and below the Galactic center (usually called the ``{\it Fermi} bubbles '')
that was found in the {\it Fermi} LAT data (e.g., Su et al. 2010)
has large spatial extent in the R CrA region.
As described below, we found residuals with a hard $\gamma$-ray spectrum
in the R CrA region, and included an additional template in equation (\ref{eq:fit_model}).

The $\gamma$-ray data in our ROIs were binned in 0.25$^{\circ}\times0.25^{\circ}$ 
pixels and fitted with equation (\ref{eq:fit_model}) in 8 logarithmically 
equally-spaced energy bins from 250 MeV to 10 GeV using a binned maximum-likelihood 
method with Poisson statistics. Low photon statistics and poor angular
resolution at low energy ($\sim$ 1.5$^{\circ}$ at 250 MeV 
under the 68\% containment radius) do not allow us to
separate components reliably. For convolution of diffuse emission with the  
instrumental response functions, we assumed an {\it E}$^{-2}$ spectrum 
and the integrated intensities were allowed to vary in each of the 8 energy
bins. Changing the fixed spectral shape index over the range from  
$-$1.5 to $-$3.0 has negligible effect on the obtained spectrum. 
Data in the R CrA region with energies above 4 GeV
are grouped in a single bin to get larger statistics.

\begin{figure}
 \begin{center}
  \includegraphics[width=80mm]{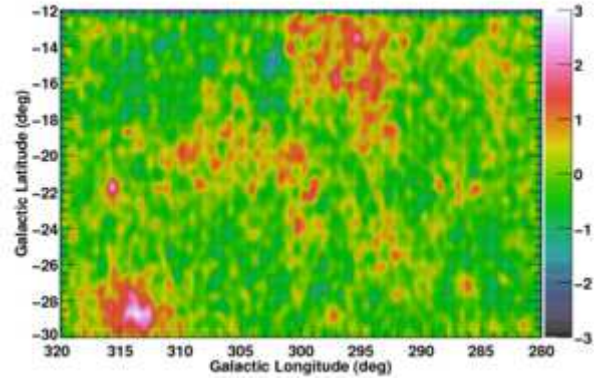}
 \end{center}
 \caption{$\gamma$-ray residual map obtained from the fit without 
 the $A{\rm v}_{\rm res}$ map in units of standard deviations above 250 MeV for the 
 Chamaeleon region smoothed with a Gaussian of $\sigma = 0.5^{\circ}$. Positive
 residual counts are seen where we have positive $A{\rm v}_{\rm res}$
 (Figure \ref{fig:Cham_gas_model_map}).}
 \label{fig:residualMap_nonDG}
\end{figure}

\begin{figure}
 \begin{center}
  \includegraphics[width=80mm]{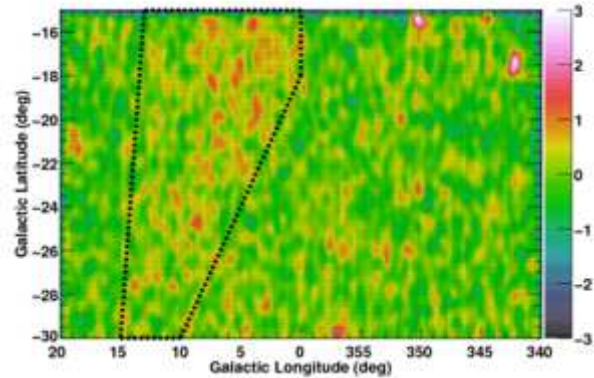}
 \end{center}
 \caption{$\gamma$-ray residual map obtained from the fit 
 in units of standard deviations above 250 MeV for the 
 R CrA region smoothed with a Gaussian of $\sigma = 0.5^{\circ}$. Dashed lines
 indicate region boundaries for the additional flat template map.}
 \label{fig:residualMap_nonFT}
\end{figure}

We started the analysis for the Chamaeleon region with point sources
detected with high significance (TS $\geq$ 100) in the 1FGL catalog. 
The normalizations for each energy bin are allowed to vary for sources
inside our ROI. We also included sources lying outside 
($\leq$ 5$^{\circ}$) ROI, with the spectral parameters fixed to those
in the 1FGL catalog. We first fitted the model to LAT data without the 
$A{\rm v}_{\rm res}$ map, and then included it and confirmed that the 
likelihood improved significantly; the test statistic, defined as 
TS $= 2({\rm ln}L_1-{\rm ln}L_0)$, where $L_0$($L_1$) is the likelihood
without(with) additional component,
is 704 for 8 more free parameters for the energy range from 250 MeV to 10 GeV. 
Figure \ref{fig:residualMap_nonDG} shows residual (data minus fitted
model) map obtained from the fit without the $A{\rm v}_{\rm res}$ map. Positive
residual counts are seen where we have positive $A{\rm v}_{\rm res}$
(Figure \ref{fig:Cham_gas_model_map}).
We thus confirmed that the positive $A{\rm v}_{\rm res}$ traces the gas
not well measured by {\HI} and CO surveys, 
and included the $A{\rm v}_{\rm res}$ map in the following analysis.
We repeated the same procedure and obtained the same conclusion for the 
Cepheus and Polaris flare region.
In the R CrA region, large residual clumps are still seen around
($0^{\circ}< l < 15^{\circ}$) and ($-30^{\circ}< b < -15^{\circ}$)
even if we included the $A{\rm v}_{\rm res}$ map
as shown by Figure \ref{fig:residualMap_nonFT},
probably due to the southern {\it Fermi} bubble.
In order to accout for these residuals, we used a flat template model map 
with the shape as shown in Figure \ref{fig:residualMap_nonFT}
with a free normalization in each energy bin.   
We note that the template map is just to accommodate the residuals 
not to investigate the {\it Fermi} bubble.
The residuals are improved significantly as shown in the bottom right panel
of Figure \ref{fig:RCrA_ana_map}, and 
the intensity of this template is too low to significantly impact on the 
local HI gas emissivity as shown in the top right panel of Figure \ref{fig:SummarySpectra}.

We then lowered the threshold for point sources down to ${\rm TS} = 50$. Although the 
fit improves in terms of the log-likelihood, the effect on the emissivities
associated with gas maps is negligible for the three regions 
(smaller than the statistical error). 
However, some point-like excesses corresponding to objects not
included in the 1FGL catalog are seen in the R CrA region 
(2FGL J1830.2-4441, 2FGL J1816.7-4942, and 2FGL J1825.1-5231) 
and Cepheus and Polaris flare region (2FGL J2022.5+7614 and 2FGL J2009.7+7225). 
These may be sources that became luminous after the 1FGL catalog was published.
We thus included those sources and confirmed that gas
emissivities are almost unaffected while the residual map becomes
flatter. We therefore adopted the model described by equation 
(\ref{eq:fit_model}) (plus a flat template for the R CrA region)
with point sources with TS $\geq$ 50 in the 1FGL
catalog (plus additional sources in the R CrA, and Cepheus 
and Polaris flare regions described above) as our baseline
model. We note that an unassociated source in the 1FGL catalog, 
1FGL J1903.8-3718c, is located on a CO core of the R CrA molecular cloud
and was recognized in the 1FGL catalog as potentially being spurious.
We thus did not include this source in our model. We confirmed that the
obtained emissivities are almost the same if we mask the region of the CO core.

\clearpage

\section{Results}
\label{sec:Results}

\begin{figure}
  \begin{minipage}{0.5\hsize}
    \includegraphics[width=80mm]{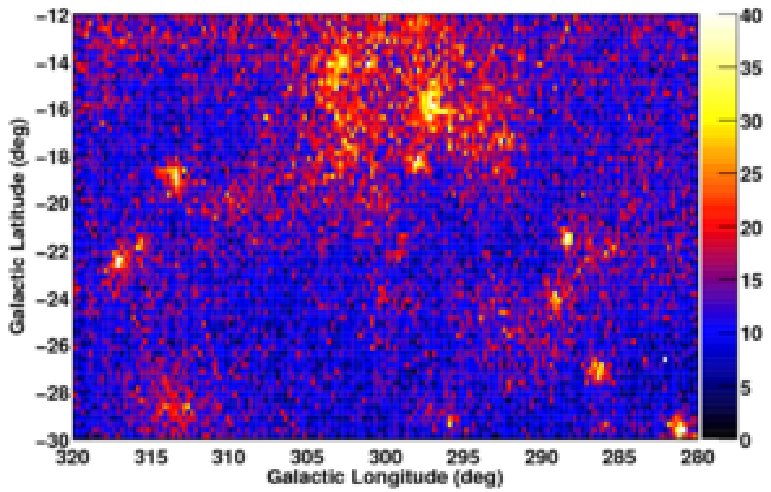}
  \end{minipage}
  \begin{minipage}{0.5\hsize}
    \includegraphics[width=80mm]{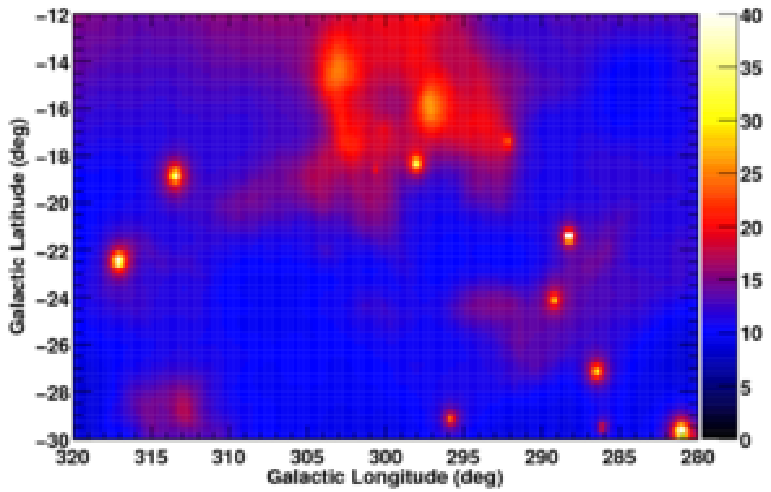}    
  \end{minipage}\\
 \begin{minipage}{0.5\hsize}
    \includegraphics[width=80mm]{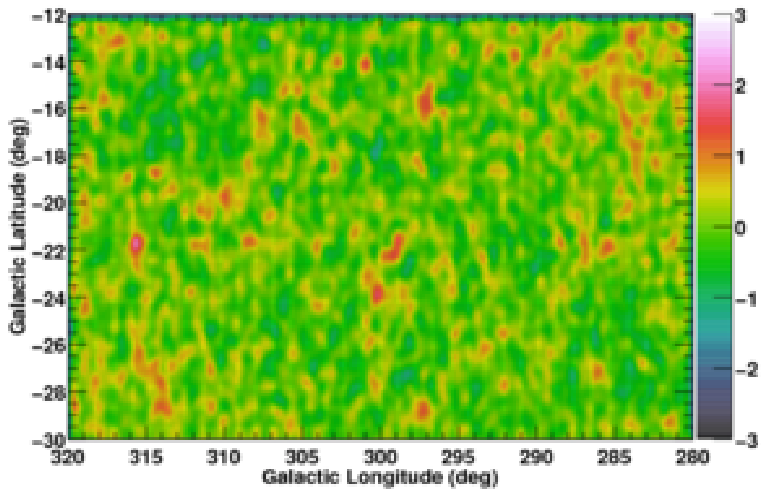}
  \end{minipage}
  \begin{minipage}{0.5\hsize}
  \end{minipage}
 \caption{Data count map (top left), fitted model count map (top right),
 and residual map (bottom left) in units of standard deviations above
 250 MeV under the assumption of {\it T}${\rm _S}$ = 125 K for the
 Chamaeleon region. The residual map has been smoothed with a Gaussian of 
 $\sigma =$ 0.5$^\circ$.}
 \label{fig:Cham_ana_map}
\end{figure}

\begin{figure}
  \begin{minipage}{0.5\hsize}
    \includegraphics[width=80mm]{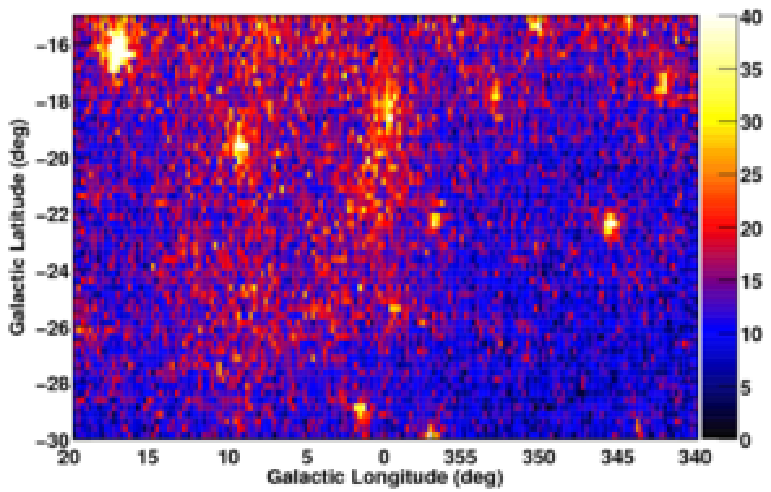}
  \end{minipage}
  \begin{minipage}{0.5\hsize}
    \includegraphics[width=80mm]{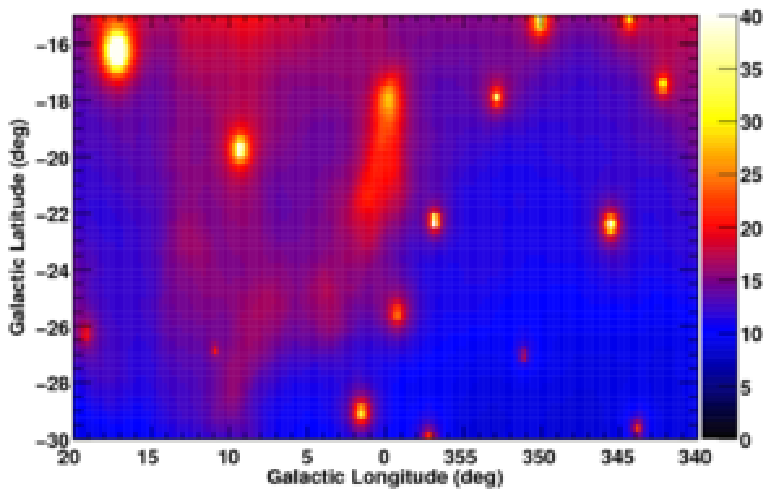}
  \end{minipage}\\
  \begin{minipage}{0.5\hsize}
    \includegraphics[width=80mm]{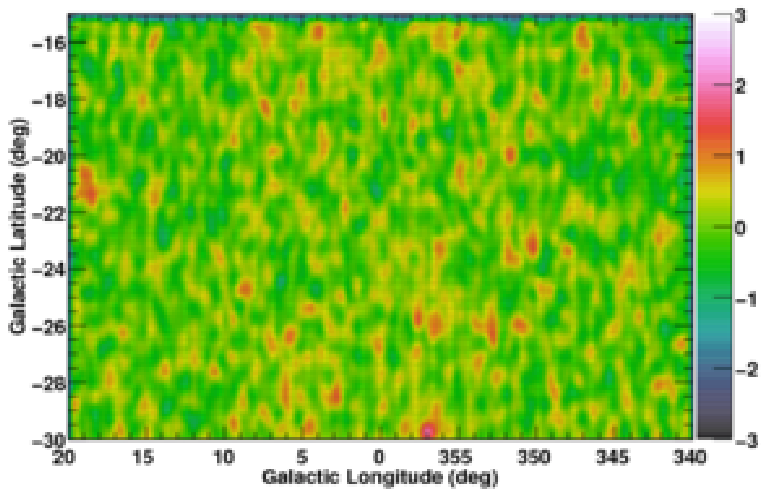}
  \end{minipage}
  \begin{minipage}{0.5\hsize}
  \end{minipage}
 \caption{The same as Figure \ref{fig:Cham_ana_map} for the R CrA region.}
 \label{fig:RCrA_ana_map}
\end{figure}

\begin{figure}
  \begin{minipage}{0.5\hsize}
    \includegraphics[width=80mm]{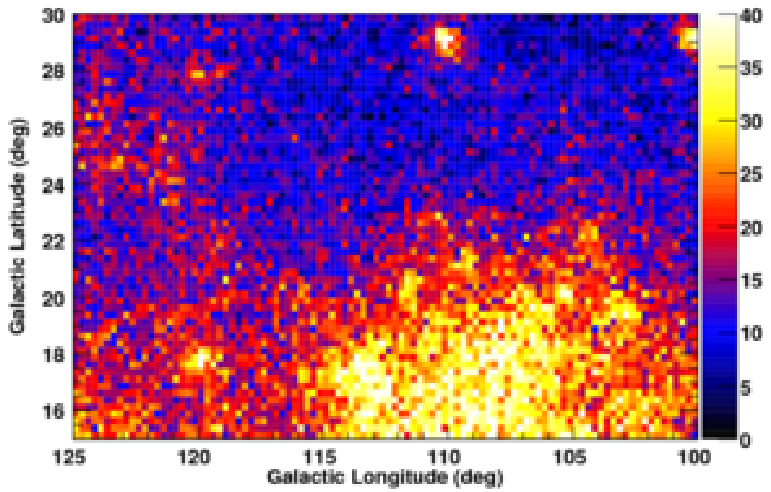}    
  \end{minipage}
  \begin{minipage}{0.5\hsize}
    \includegraphics[width=80mm]{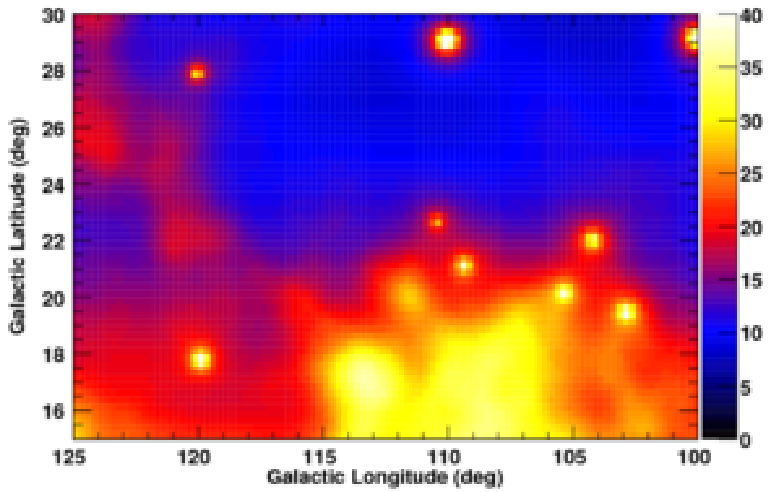}
  \end{minipage}\\
  \begin{minipage}{0.5\hsize}
    \includegraphics[width=80mm]{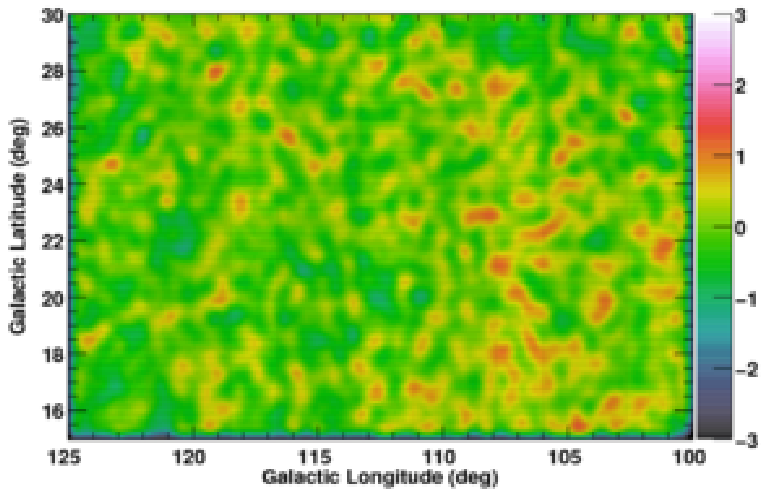}
  \end{minipage}
  \begin{minipage}{0.5\hsize}
  \end{minipage}
 \caption{The same as Figure \ref{fig:Cham_ana_map} for the Cepheus and Polaris flare region.}
 \label{fig:CePo_ana_map}
\end{figure}

\begin{figure}
  \begin{minipage}{0.5\hsize}
    \includegraphics[width=80mm]{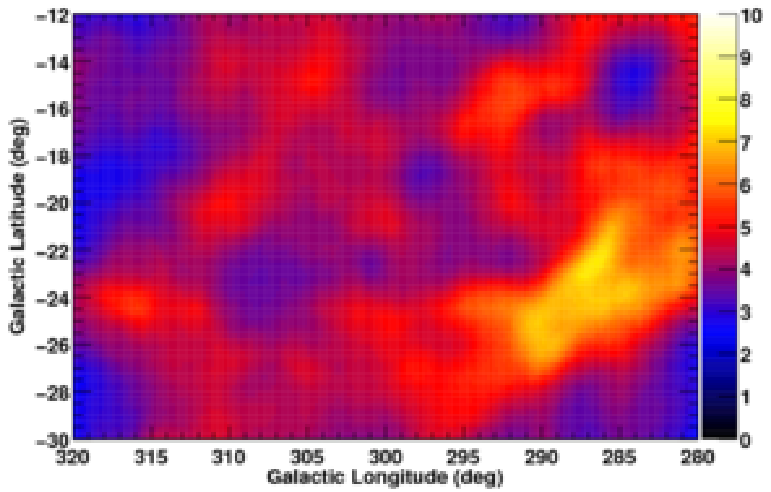}
    \label{fig:HI_map}
  \end{minipage}
  \begin{minipage}{0.5\hsize}
    \includegraphics[width=80mm]{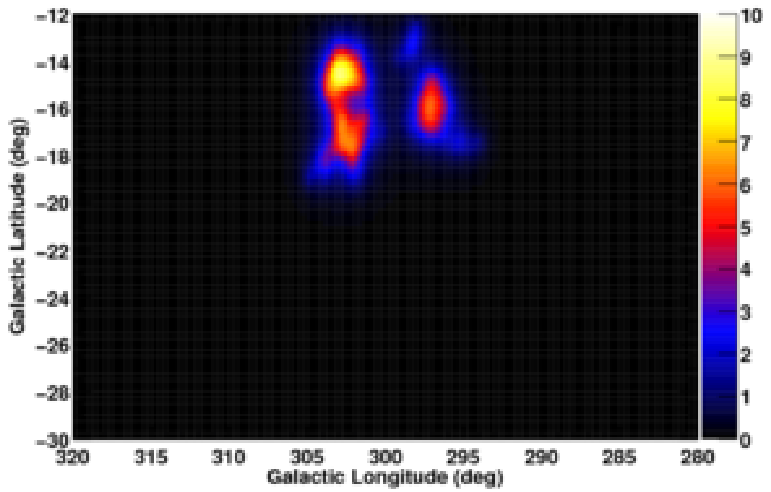}
    \label{fig:CO_map}
  \end{minipage}\\
  \begin{minipage}{0.5\hsize}
    \includegraphics[width=80mm]{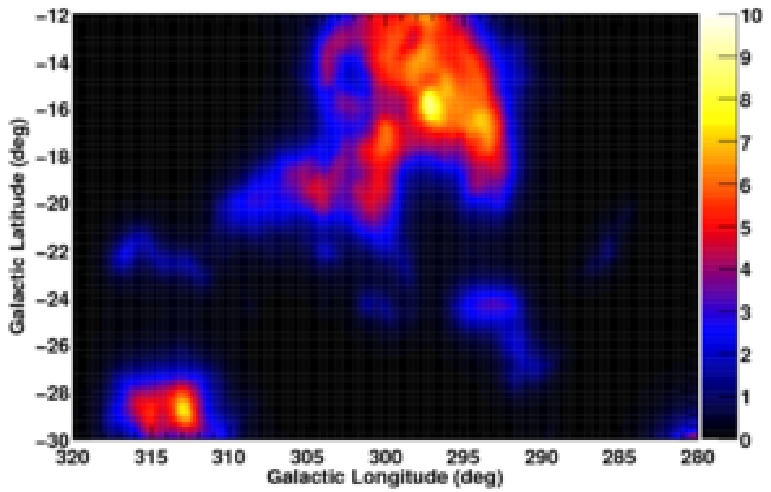}
    \label{fig:DG_map}
  \end{minipage}
  \begin{minipage}{0.5\hsize}
  \end{minipage}
 \caption{Fitted model count maps above 250 MeV for the Chamaeleon
 region; {\HI} component (top left), CO component (top right) and 
 $A{\rm v}_{\rm res}$ component (bottom left).}
 \label{fig:model_maps}
\end{figure}

Figures \ref{fig:Cham_ana_map}, \ref{fig:RCrA_ana_map}, and \ref{fig:CePo_ana_map} 
show the $\gamma$-ray data count maps, fitted model count maps and the residual maps
for the Chamaeleon, R CrA, and Cepheus and Polaris flare regions, respectively, 
in which the residuals are expressed in units of approximate 
standard deviations (square root of the model counts). 
The residual maps show no conspicuous structures, indicating
that our model reasonably reproduces the data, particularly the diffuse
emission. For illustrative purposes, we present the fitted model count
maps for the Chamaeleon region decomposed into each gas component in 
Figure \ref{fig:model_maps}, in which the $\gamma$-ray emission from 
{\HI}, that from the molecular gas traced by $W_{\rm CO}$,  
and that inferred from the $A{\rm v}_{\rm res}$ map are shown. 
Although the distribution of {\it N}({\HI}) is rather uniform in our
ROI, it exhibits some structures and allows us to derive the emissivity
of the {\HI} gas, which is proportional to the flux of
ambient CRs. The distribution of $W_{\rm CO}$ is highly structured and 
is concentrated in the longitude range from 295$^{\circ}$ to
305$^{\circ}$ and in the latitude range from $-12^{\circ}$ to
$-20^{\circ}$. The gas traced by $A{\rm v}_{\rm res}$ lies at the
interface between the {\HI} component (atomic gas) and the
CO component (molecular gas), and has a mass (proportional to the $\gamma$-ray
counts) comparable to or larger than that of the molecular gas traced by $W_{\rm CO}$.

\begin{figure}
 \begin{minipage}{0.5\hsize}
    \includegraphics[width=80mm]{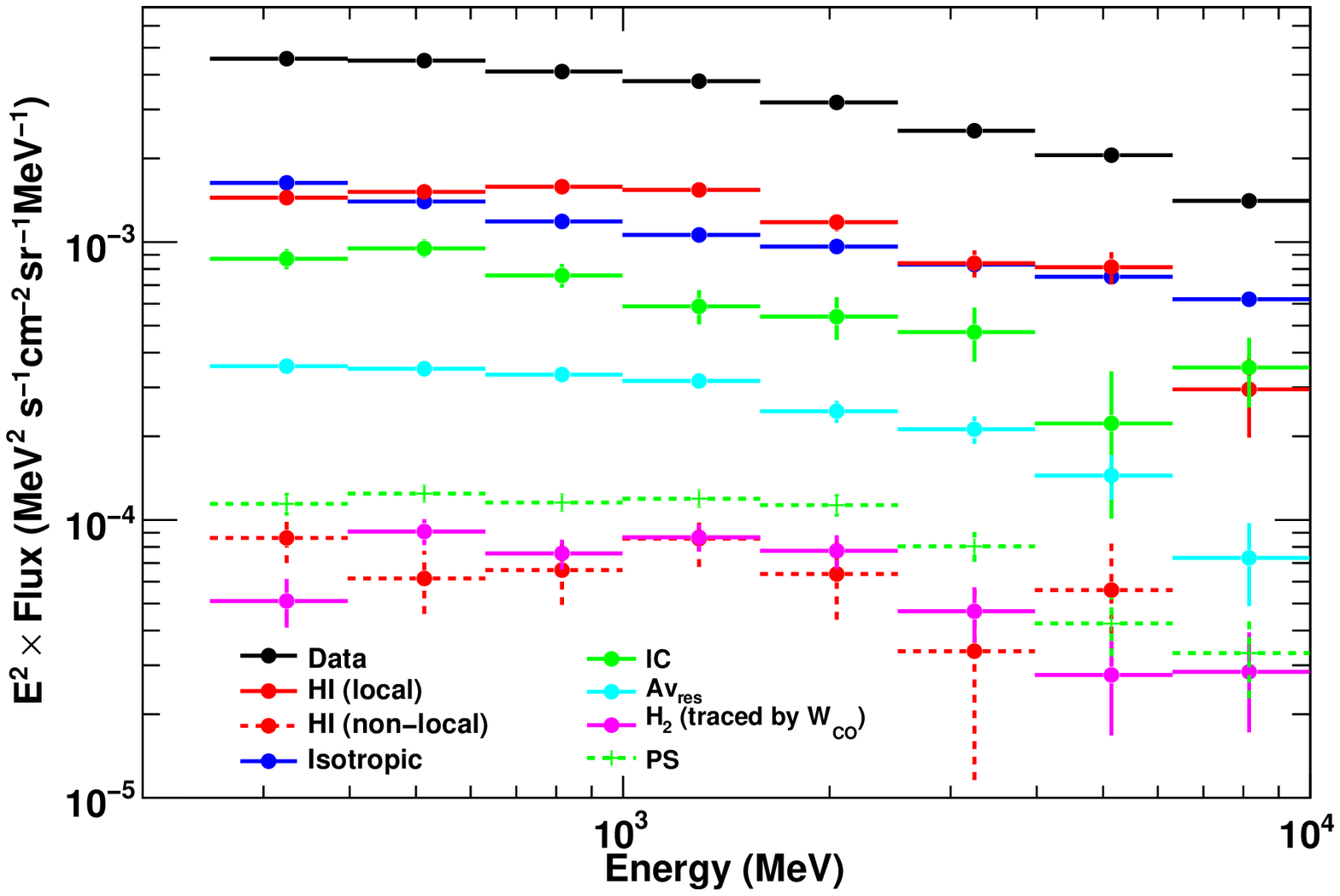}
    \label{fig:ChamSummarySpectra}
  \end{minipage} 
 \begin{minipage}{0.5\hsize}
    \includegraphics[width=80mm]{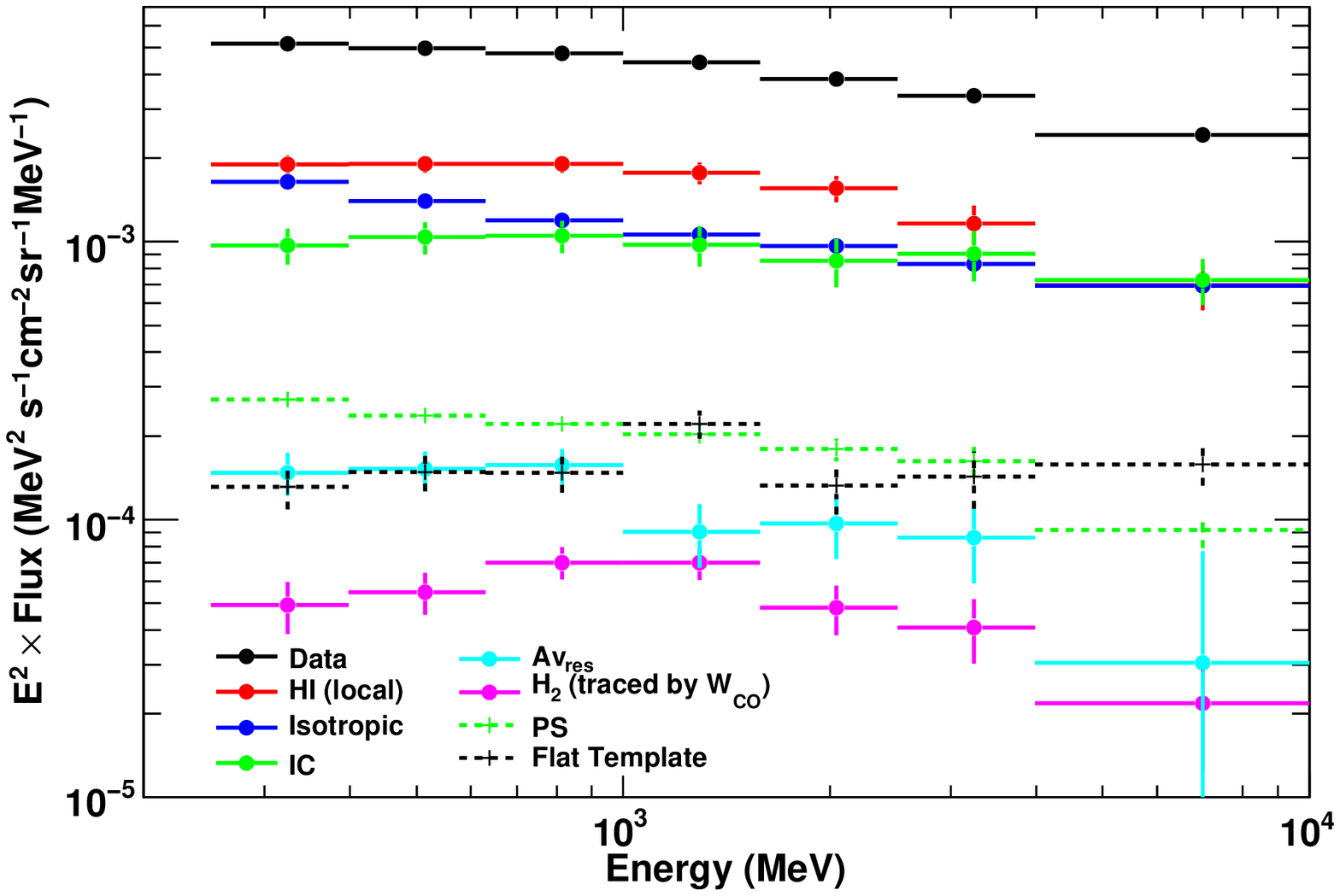}
    \label{fig:RCrASummarySpectra}
  \end{minipage}\\
  \begin{minipage}{0.5\hsize}
    \includegraphics[width=80mm]{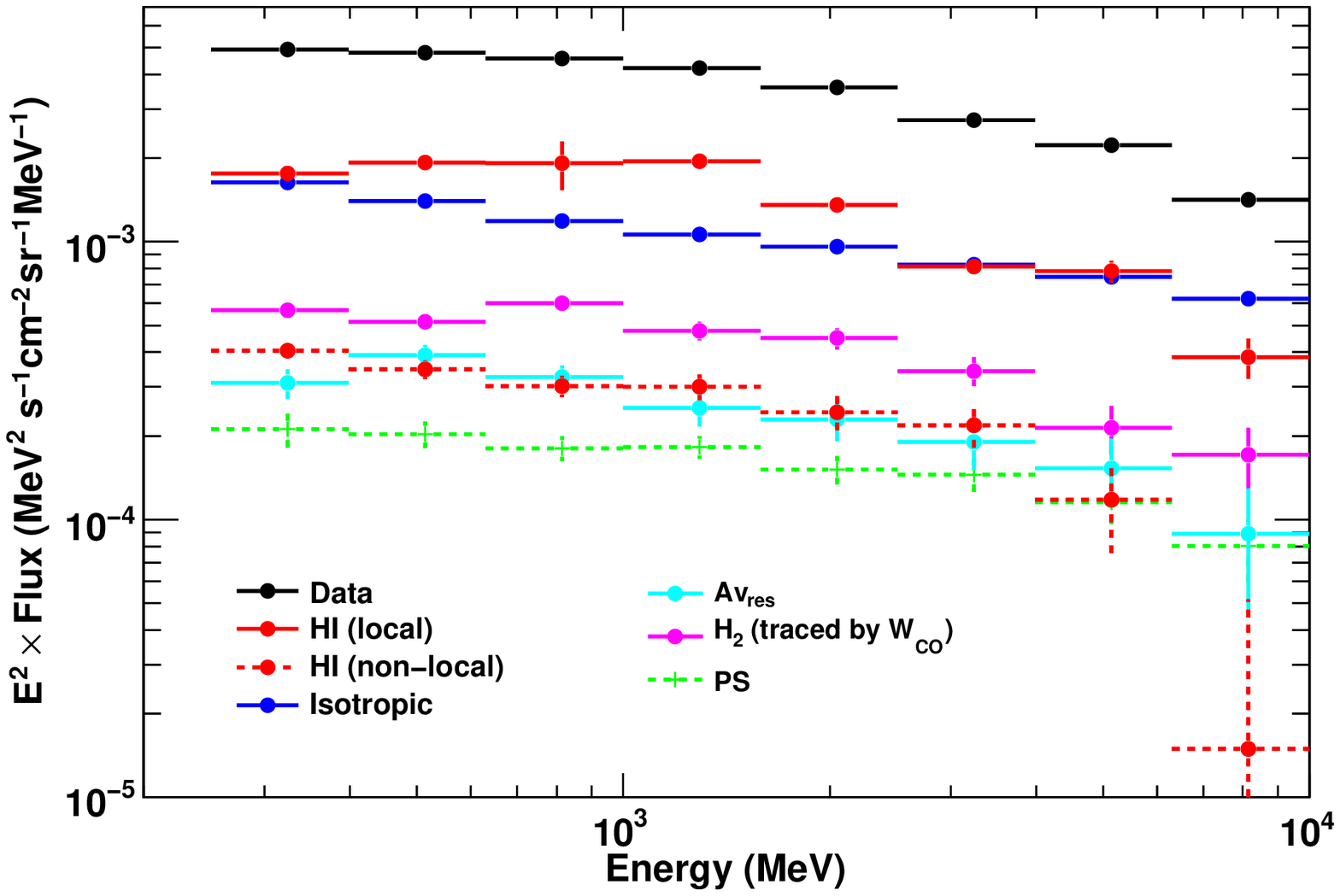}
    \label{fig:CePoSummarySpectra}
  \end{minipage} 
 \begin{minipage}{0.5\hsize}
  \end{minipage}
 \caption{$\gamma$-ray spectra of each component (each gas phase, IC,
 isotropic component, point sources, and flat template model for the 
 R CrA region) for the Chamaeleon (top left), R CrA (top right), 
 and Cepheus and Polaris flare region (bottom left). The {\HI} (non-local) 
 for the R CrA region and the IC component for the Cepheus and Polaris flare 
 are not shown here since they are negligible.}
 \label{fig:SummarySpectra}
\end{figure}

Figure \ref{fig:SummarySpectra} shows the fitted spectra for each component. 
Although the contributions from IC and isotropic components are large, 
the spectra of each gas component are reliably constrained due to 
their characteristic spatial distributions as shown in Figures 
\ref{fig:Cham_gas_model_map}, \ref{fig:RCrA_gas_model_map}, and \ref{fig:CePo_gas_model_map}.
The hard spectra of the IC term and the flat template model component of the R CrA region are
likely to be due to the southern {\it Fermi} bubble.

The fit results are summarized in Tables \ref{table:fit_results_Cham},
\ref{table:fit_results_R_CrA}, and \ref{table:fit_results_CePo} 
for the case of {\it T}${\rm _S}$ = 125 K
in the Chamaeleon, R CrA, and Cepheus and Polaris flare regions, respectively. 
The integrated {\HI} emissivity for the Chamaeleon region above
250 MeV is (5.9 $\pm$ 0.1$_{\rm stat}$ $^{+0.9}_{-1.0}$$_{\rm sys}$) 
$\times$ 10$^{-27}$ photons s$^{-1}$ sr$^{-1}$ H-atom$^{-1}$,
and those of the R CrA, and Cepheus and Polaris flare regions are 
(10.2 $\pm$ 0.4$_{\rm stat}$ $^{+1.2}_{-1.7}$$_{\rm sys}$) 
$\times$ 10$^{-27}$ photons s$^{-1}$ sr$^{-1}$ H-atom$^{-1}$
and (9.1 $\pm$ 0.3$_{\rm stat}$ $^{+1.5}_{-0.6}$$_{\rm sys}$) 
$\times$ 10$^{-27}$ photons s$^{-1}$ sr$^{-1}$ H-atom$^{-1}$, respectively.
(See below for the evaluation of the systematic uncertainty from the sky model.)   

\begin{table}[t]
 \caption{\normalsize{ Gas emissivities in the Chamaeleon regions with 
  their statistical uncertainties (assuming {\it T}${\rm _S}$ = 125 K for the {\HI} 
  maps preparation).}}
 \label{table:fit_results_Cham}
  \begin{center}
   \begin{tabular}{cccc} \hline\hline
   \makebox[5em][c]{Energy} &
   \makebox[10em][c]{$q_{{\rm HI},1}$} &
   \makebox[10em][c]{$q_{\rm CO}$} &
   \makebox[10em][c]{$q_{\rm Av}$} \\ 
   (GeV)     &               &                &               \\ \hline
   0.25-0.40 & 2.3$\pm$0.1 & 0.30$\pm$0.06  & 0.57$\pm$0.03 \\
   0.40-0.63 & 1.50$\pm$0.06 & 0.33$\pm$0.04  & 0.34$\pm$0.02 \\
   0.63-1.00 & 0.99$\pm$0.04 & 0.17$\pm$0.02  & 0.21$\pm$0.01 \\
   1.00-1.58 & 0.61$\pm$0.03 & 0.13$\pm$0.01  & 0.125$\pm$0.008 \\
   1.58-2.51 & 0.29$\pm$0.02 & 0.069$\pm$0.009 & 0.060$\pm$0.006 \\
   2.51-3.98 & 0.13$\pm$0.01 & 0.027$\pm$0.006 & 0.033$\pm$0.004\\
   3.98-6.31 & 0.08$\pm$0.01 & 0.010$\pm$0.004  & 0.014$\pm$0.003 \\
   6.31-10.00 & 0.019$\pm$0.006 & 0.007$\pm$0.003  & 0.005$\pm$0.002 \\
   total & 5.9$\pm$0.1 & 1.04$\pm$0.08 & 1.36$\pm$0.04 \\ \hline
   \multicolumn{4}{c}{\scriptsize{{\bf Notes.} units; $q_{{\rm HI},1}$(10$^{-27}$ s$^{-1}$
   sr$^{-1}$), $q_{\rm CO}$(10$^{-6}$ cm$^{-2}$ s$^{-1}$ sr$^{-1}$ (K km 
   s$^{-1}$)$^{-1}$), $q_{\rm Av}$(10$^{-5}$ cm$^{-2}$ s$^{-1}$
   sr$^{-1}$ mag$^{-1}$)}} \\ 
   \end{tabular}
  \end{center}
\end{table}

\begin{table}
 \caption{\normalsize{The same as Table \ref{table:fit_results_Cham} 
 for the R CrA region.}}
 \label{table:fit_results_R_CrA}
  \begin{center}
   \begin{tabular}{cccc}\hline\hline
   \makebox[5em][c]{Energy} &
   \makebox[10em][c]{$q_{\rm HI}$} &
   \makebox[10em][c]{$q_{\rm CO}$} &
   \makebox[10em][c]{$q_{\rm Av}$} \\
   (GeV)     &                &                & \\ \hline
   0.25-0.40 & 4.1$\pm$0.3  & 0.7$\pm$0.1  & 1.0$\pm$0.2 \\
   0.40-0.63 & 2.6$\pm$0.2  & 0.45$\pm$0.08  & 0.6$\pm$0.1 \\
   0.63-1.00 & 1.6$\pm$0.1  & 0.37$\pm$0.05  & 0.41$\pm$0.06 \\
   1.00-1.58 & 0.96$\pm$0.09  & 0.23$\pm$0.03  & 0.15$\pm$0.04 \\
   1.58-2.51 & 0.53$\pm$0.06  & 0.10$\pm$0.02  & 0.10$\pm$0.03 \\
   2.51-3.98 & 0.25$\pm$0.04  & 0.05$\pm$0.01  & 0.06$\pm$0.02 \\
   3.98-10.00 & 0.14$\pm$0.03 & 0.019$\pm$0.008  & 0.008$\pm$0.001 \\ 
   total      & 10.2$\pm$0.4  & 1.9$\pm$0.2 & 2.3$\pm$0.2 \\ \hline
   \multicolumn{4}{c}{\scriptsize{{\bf Notes.} units; $q_{\rm HI}$(10$^{-27}$ s$^{-1}$
    sr$^{-1}$), $q_{\rm CO}$(10$^{-6}$ cm$^{-2}$ s$^{-1}$ sr$^{-1}$ (K
    km s$^{-1}$)$^{-1}$), $q_{\rm Av}$(10$^{-5}$ cm$^{-2}$ s$^{-1}$
    sr$^{-1}$ mag$^{-1}$)}} 
   \end{tabular}
  \end{center}
\end{table}

\begin{table}
 \caption{\normalsize{The same as Table \ref{table:fit_results_Cham}
 for the Cepheus and Polaris flare region.}}
 \label{table:fit_results_CePo}
  \begin{center}
   \begin{tabular}{cccc}\hline\hline
   \makebox[5em][c]{Energy} &
   \makebox[10em][c]{$q_{\rm HI}$} &
   \makebox[10em][c]{$q_{\rm CO}$} &
   \makebox[10em][c]{$q_{\rm Av}$} \\
   (GeV)     &                &                & \\ \hline
   0.25-0.40 & 3.5$\pm$0.1  & 0.52$\pm$0.04  & 0.53$\pm$0.06 \\
   0.40-0.63 & 2.37$\pm$0.06  & 0.29$\pm$0.02  & 0.41$\pm$0.04 \\
   0.63-1.00 & 1.6$\pm$0.3  & 0.21$\pm$0.01  & 0.22$\pm$0.02 \\
   1.00-1.58 & 0.97$\pm$0.03  & 0.105$\pm$0.008  & 0.11$\pm$0.02 \\
   1.58-2.51 & 0.48$\pm$0.02  & 0.061$\pm$0.005  & 0.06$\pm$0.01 \\
   2.51-3.98 & 0.20$\pm$0.01  & 0.029$\pm$0.003  & 0.030$\pm$0.006 \\
   3.98-6.31 & 0.11$\pm$0.01  & 0.012$\pm$0.002  & 0.016$\pm$0.004 \\
   6.31-10.00 & 0.035$\pm$0.006 & 0.006$\pm$0.001  & 0.006$\pm$0.003 \\
   total      & 9.2$\pm$0.3  & 1.23$\pm$0.05 & 1.38$\pm$0.08 \\ \hline
   \multicolumn{4}{c}{\scriptsize{{\bf Notes.} units; $q_{\rm HI}$(10$^{-27}$ s$^{-1}$
    sr$^{-1}$), $q_{\rm CO}$(10$^{-6}$ cm$^{-2}$ s$^{-1}$ sr$^{-1}$ (K
    km s$^{-1}$)$^{-1}$), $q_{\rm Av}$(10$^{-5}$ cm$^{-2}$ s$^{-1}$
    sr$^{-1}$ mag$^{-1}$)}} 
   \end{tabular}
  \end{center}
\end{table}

\begin{figure}
  \begin{minipage}{0.5\hsize}
    \includegraphics[width=80mm]{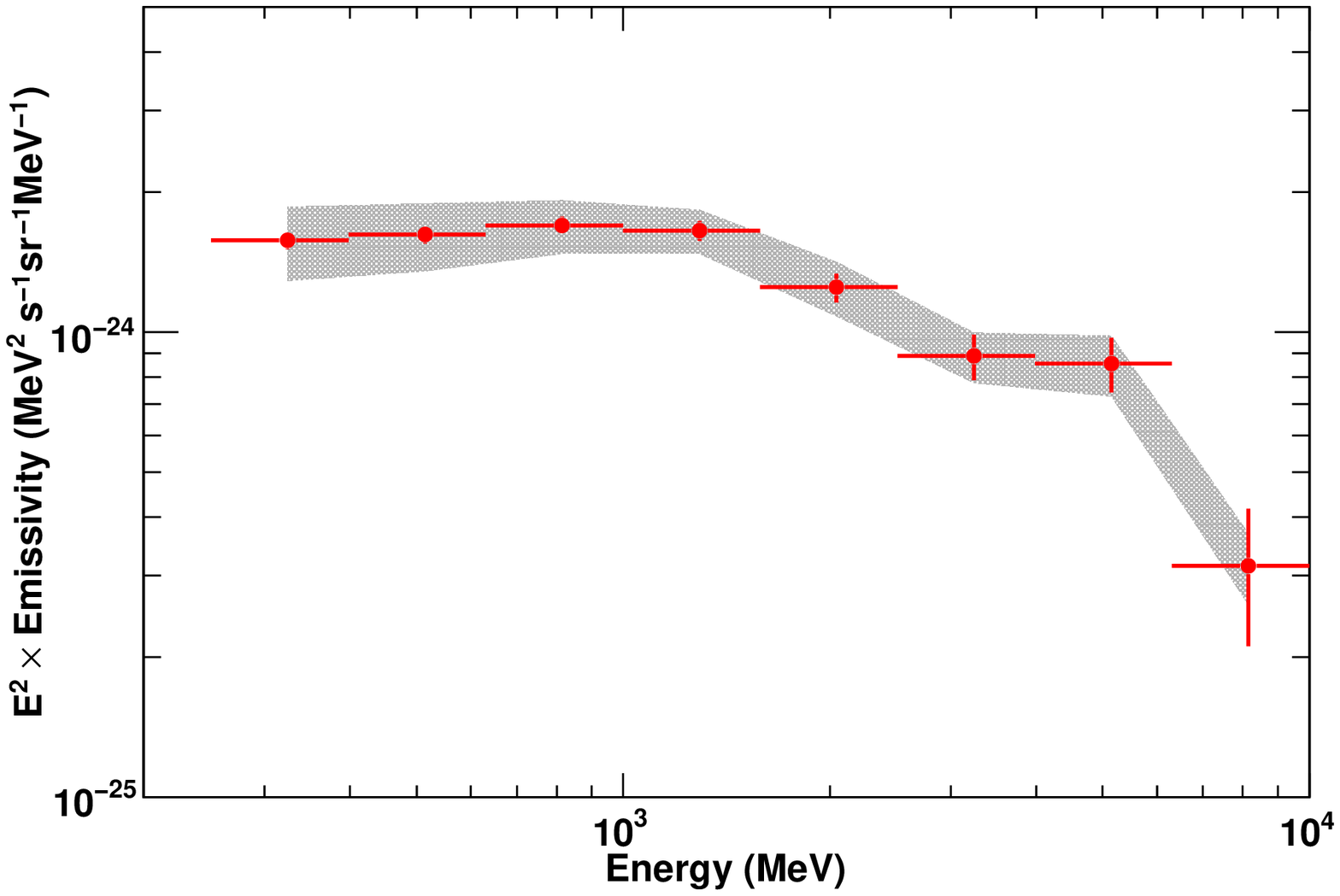}
    \label{fig:Cham_HI_emissivity}
  \end{minipage}
  \begin{minipage}{0.5\hsize}
    \includegraphics[width=80mm]{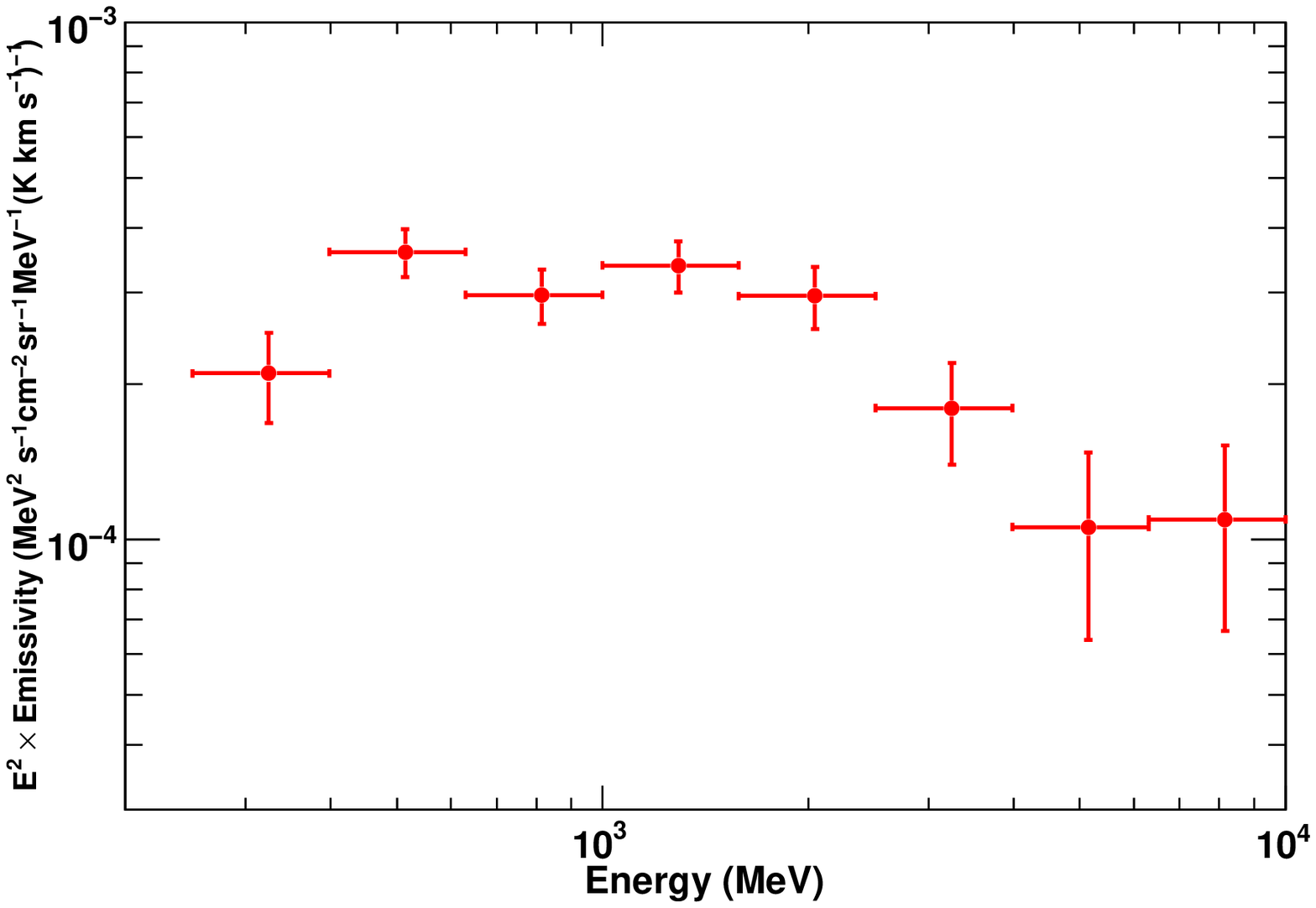}
    \label{fig:Cham_CO_emissivity}
  \end{minipage}\\ 
  \begin{minipage}{0.5\hsize}
    \includegraphics[width=80mm]{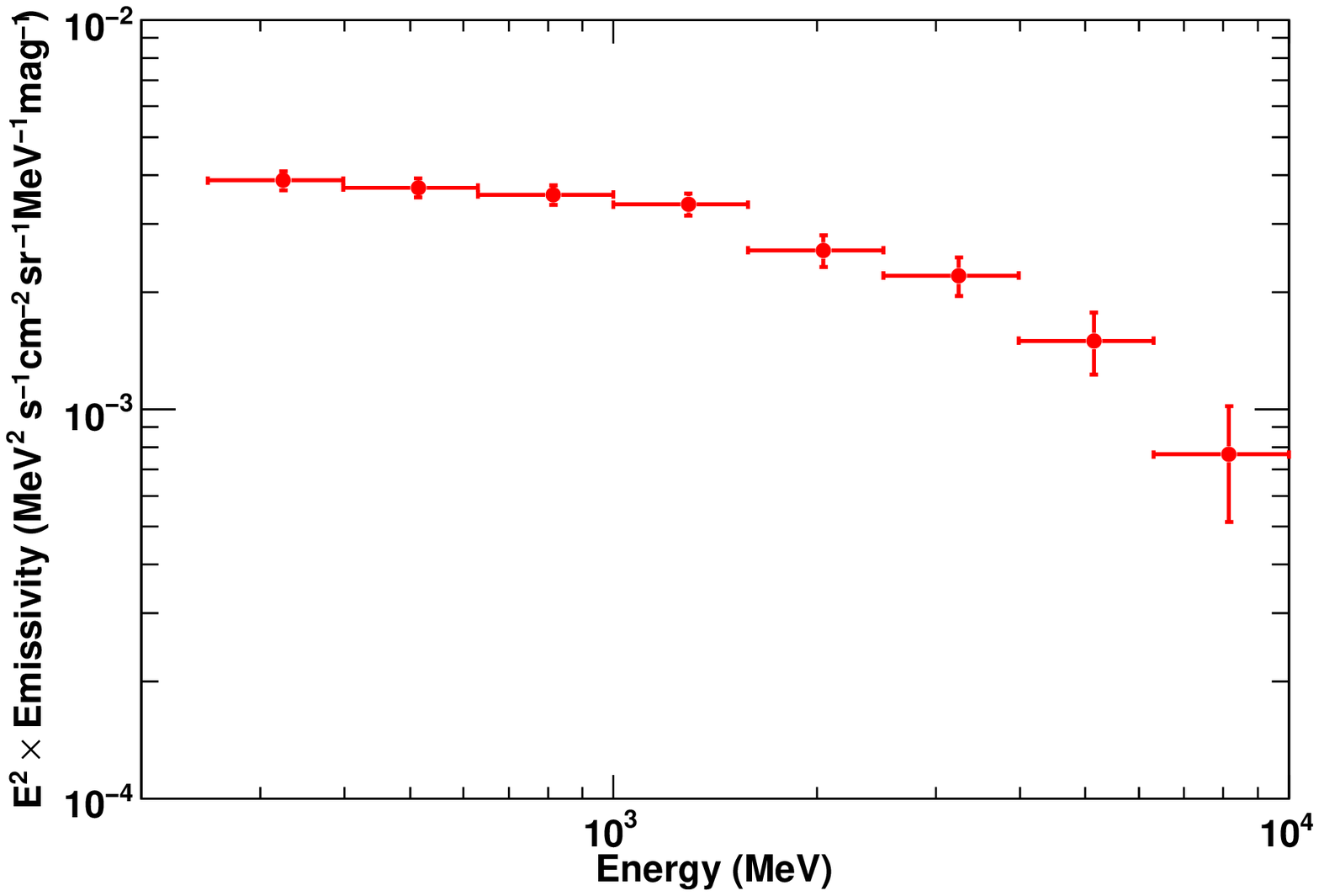}
    \label{fig:Cham_DG_emissivity}
  \end{minipage}
   \begin{minipage}{0.5\hsize}
   \end{minipage}
  \caption{Emissivity spectrum of the local {\HI} gas (top left),
  that per $W_{\rm CO}$ unit (top right) and that per unit 
  $A{\rm v}_{\rm res}$ magnitude (bottom left) of the Chamaeleon region. 
  The shaded area shows systematic uncertainties for {\HI} 
  (see text for details).}
 \label{fig:Cham_emissivity}
\end{figure}

\begin{figure}
  \begin{minipage}{0.5\hsize}
    \includegraphics[width=80mm]{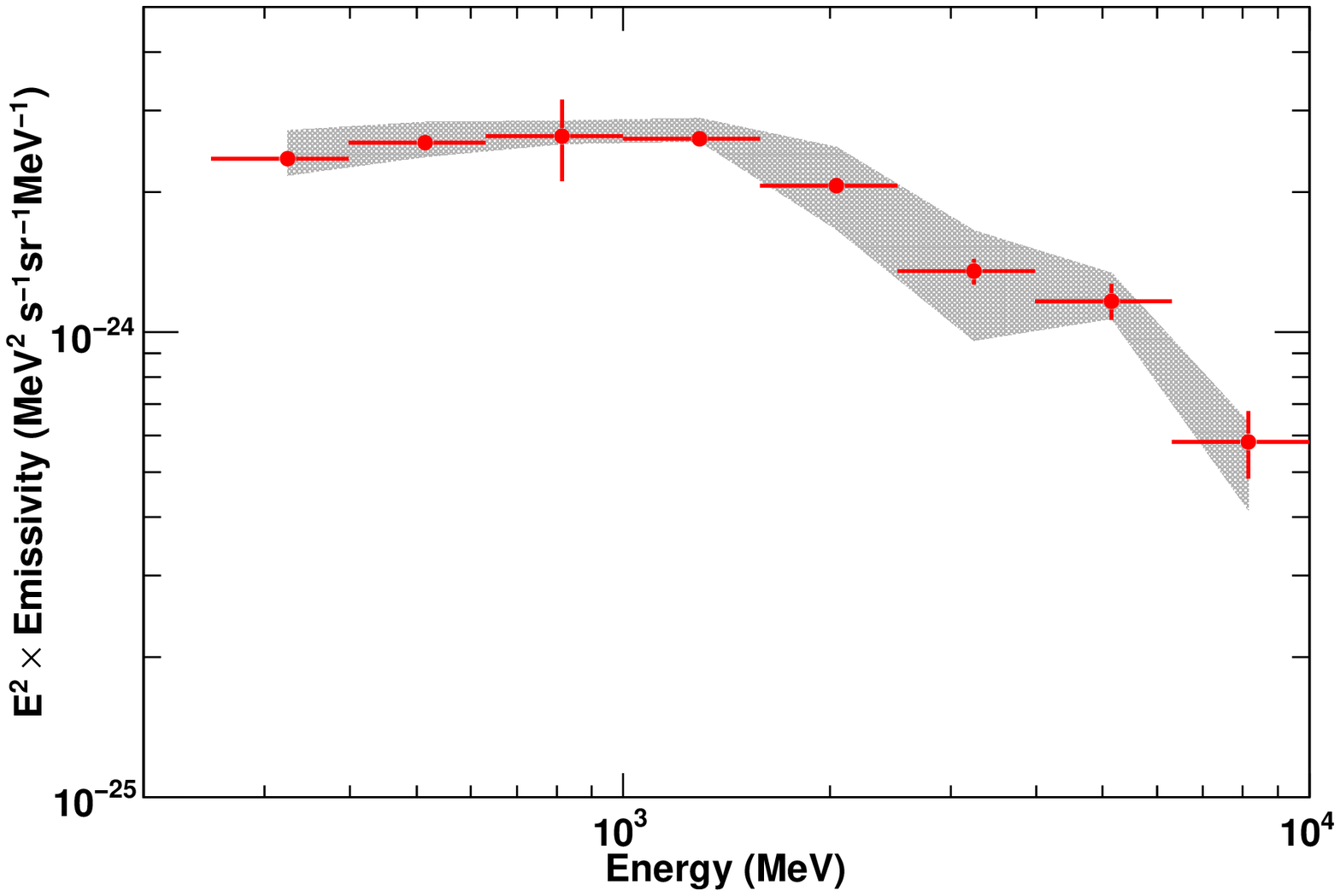}
   \end{minipage} 
  \begin{minipage}{0.5\hsize}
    \includegraphics[width=80mm]{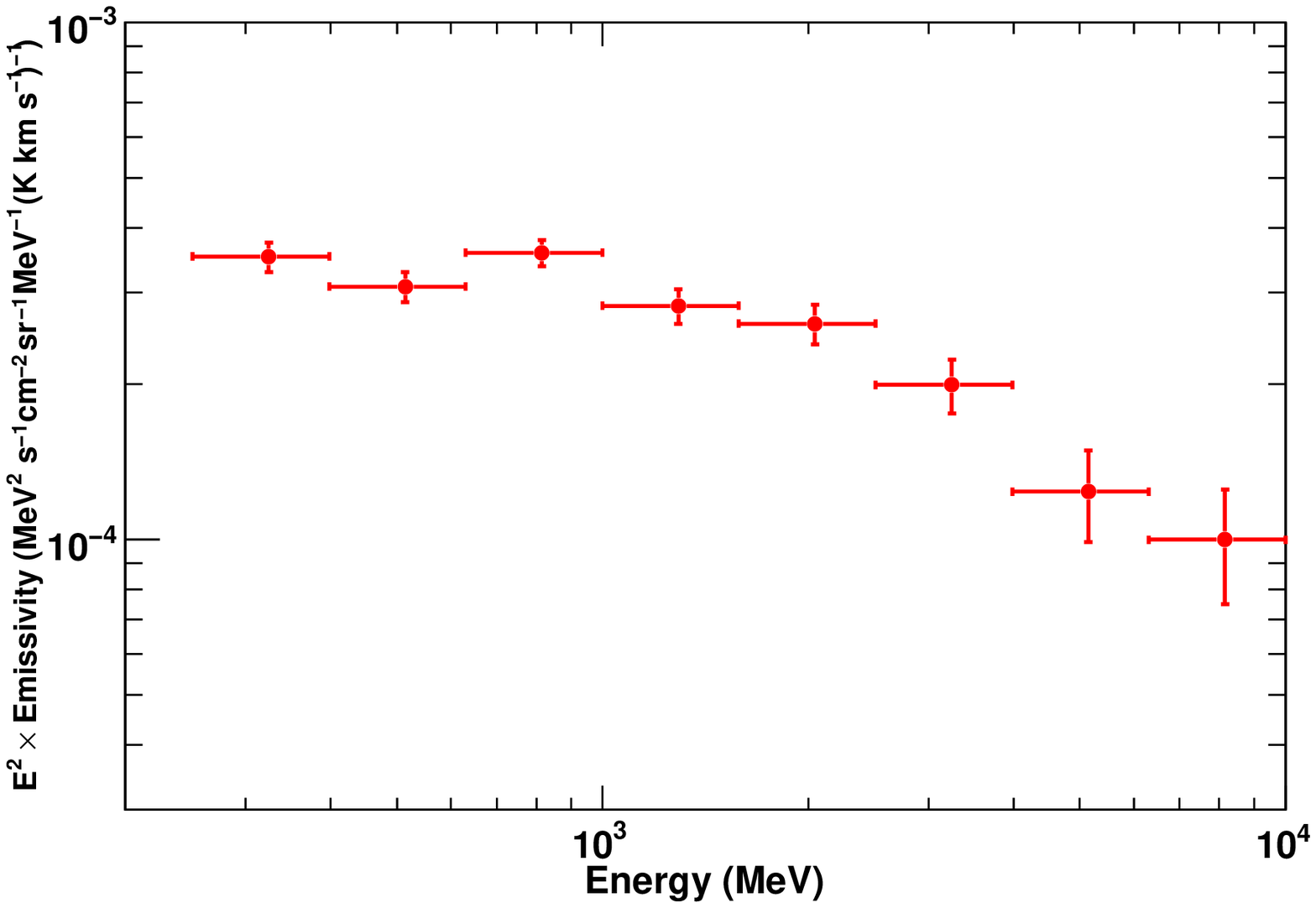}    
  \end{minipage}\\
  \begin{minipage}{0.5\hsize}
    \includegraphics[width=80mm]{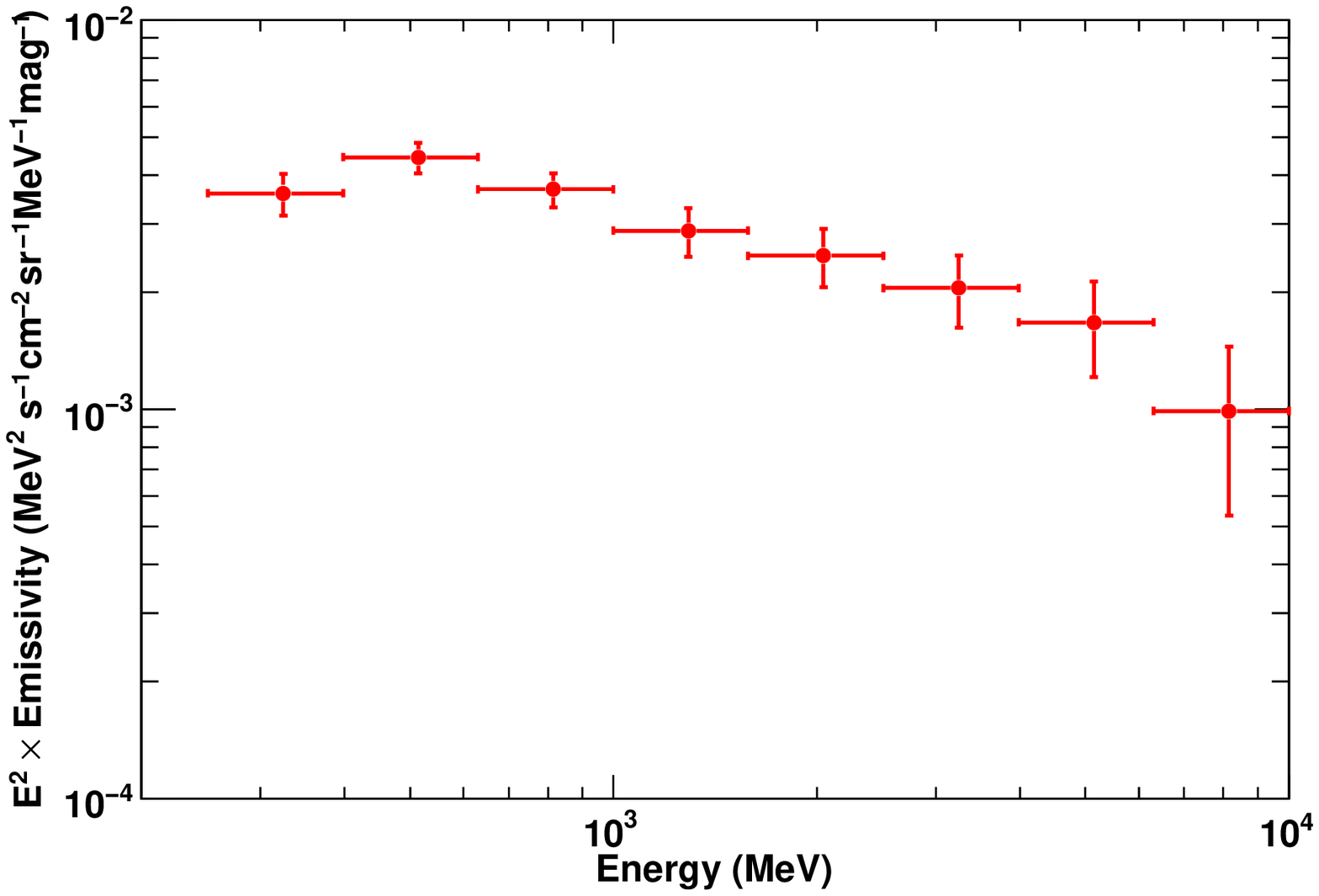}
  \end{minipage}
  \begin{minipage}{0.5\hsize}
  \end{minipage}
  \caption{The same as Figure \ref{fig:Cham_emissivity} for the Cepheus and Polaris flare region.}
 \label{fig:CePo_emissivity}
\end{figure}

Figure \ref{fig:Cham_emissivity} shows the emissivity spectra of each gas component in
the Chamaeleon region under the assumption of {\it T}${\rm _S}$ = 125 K.
In order to examine the systematic uncertainty due to the optical depth
correction, we also tried to fit the data with maps obtained by assuming 
{\it T}${\rm _S}$ = 100 K and under the approximation that the gas is
optically-thin. We evaluated the uncertainty of the isotropic component to be $\pm$10\% 
by comparing the model we adopted and those derived in
other LAT studies of mid-latitude regions (Abdo et al. 2009b and Abdo et al. 2009c). 
We thus reran the analysis described in Section \ref{sec:Analysis_procedure} 
assuming a 10\% higher and lower intensity for the fixed isotropic component. 
We also investigated the effect on the systematic uncertainty due to the
IC component by using different IC model maps, as described in Section \ref{sec:IC_Iso_PS}.
The effects of the uncertainty of the {\it T}${\rm _S}$, isotropic component, and IC models
are quite comparable, therefore we added them.
The obtained systematic uncertainty is comparable to or slightly larger than 
the statistical error as shown by Figure \ref{fig:Cham_emissivity} for $q_{\rm HI}$ (top left panel).
On the other hand, the systematic uncertainty of the $q_{\rm CO}$ and 
$q_{\rm Av}$ is smaller than the statistical error and is not shown in this figure for clarity. 
The obtained spectra in the R CrA region and the Cepheus and Polaris flare region 
are summarized in Figures \ref{fig:R_CrA_emissivity} and 
\ref{fig:CePo_emissivity}, respectively. We performed the same procedure and evaluated 
the systematic uncertainties due to the {\it T}${\rm _S}$, 
isotropic component, and IC models as a shaded area in the figures.

Evaluation of the systematic uncertainty is crucial for this study.
We therefore performed two more tests.

\begin{enumerate}
\item We modified the longitude range from $-20^{\circ}<l<20^{\circ}$ to 
 $-30^{\circ}<l<10^{\circ}$ for the R CrA region, and found that the obtained
 {\HI} emissivity was lower by $\sim$ 10 \%. This is due to a coupling between 
 the {\HI} and IC components, and take this effect on the {\HI} emissivity 
 into account in the evaluation of the overall systematic uncertainty.
\item The prominent {\HI} cloud at $280^{\circ}<l<295^{\circ}$ and 
 $-28^{\circ}<b<-20^{\circ}$ is not spatially associated with the Chamaeleon 
 molecular cloud, and it may not be physically related to the 
 molecular cloud. We masked this region and found that the {\HI} emissivity is 
 almost unaffected. 
\end{enumerate} 

The resultant peak-to-peak uncertainty of the local {\HI} emissivity is less 
than $\sim$20 \% across the energy range for three regions investigated.

\clearpage

\begin{figure}
  \begin{minipage}{0.5\hsize}
    \includegraphics[width=80mm]{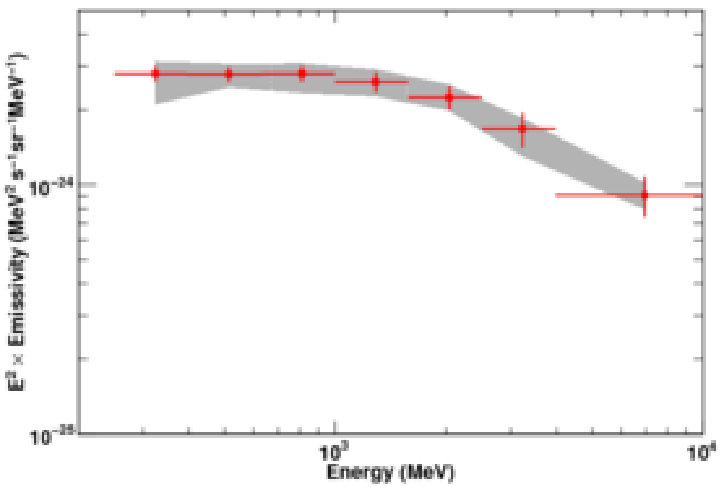}
    \label{fig:RCrA_HI_emissivity}
  \end{minipage}
  \begin{minipage}{0.5\hsize}
    \includegraphics[width=80mm]{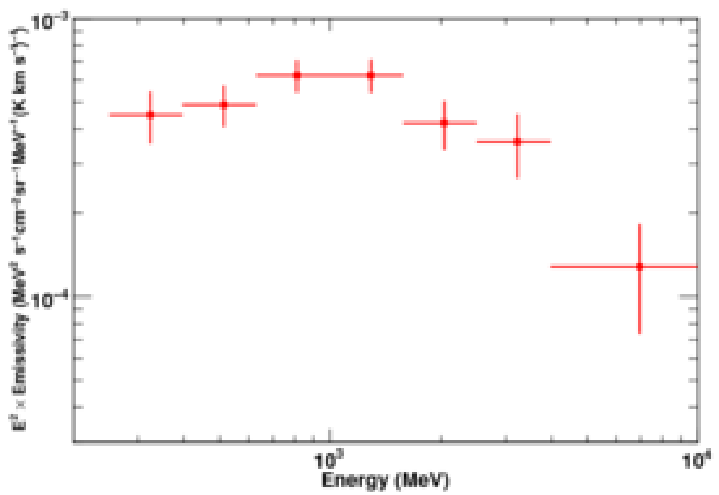}
    \label{fig:RCrA_CO_emissivity}
  \end{minipage}\\ 
  \begin{minipage}{0.5\hsize}
    \includegraphics[width=80mm]{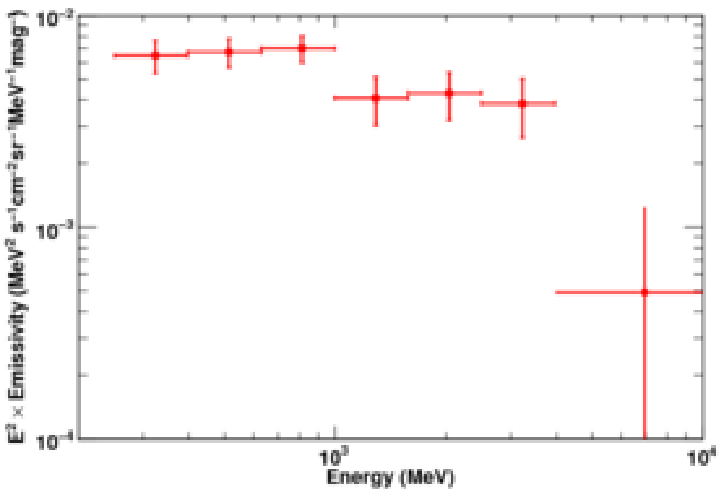}
    \label{fig:RCrA_DG_emissivity}
  \end{minipage}
   \begin{minipage}{0.5\hsize}
   \end{minipage}
  \caption{The same as Figure \ref{fig:Cham_emissivity} for the R CrA region.}
 \label{fig:RCrA_emissivity}
\end{figure}


\clearpage

\clearpage

\section{Discussion}
\label{sec:Discussion}

\subsection{CR Density and Spectrum Close to the Solar System}

\begin{figure}
 \begin{center}
  \includegraphics[width=80mm]{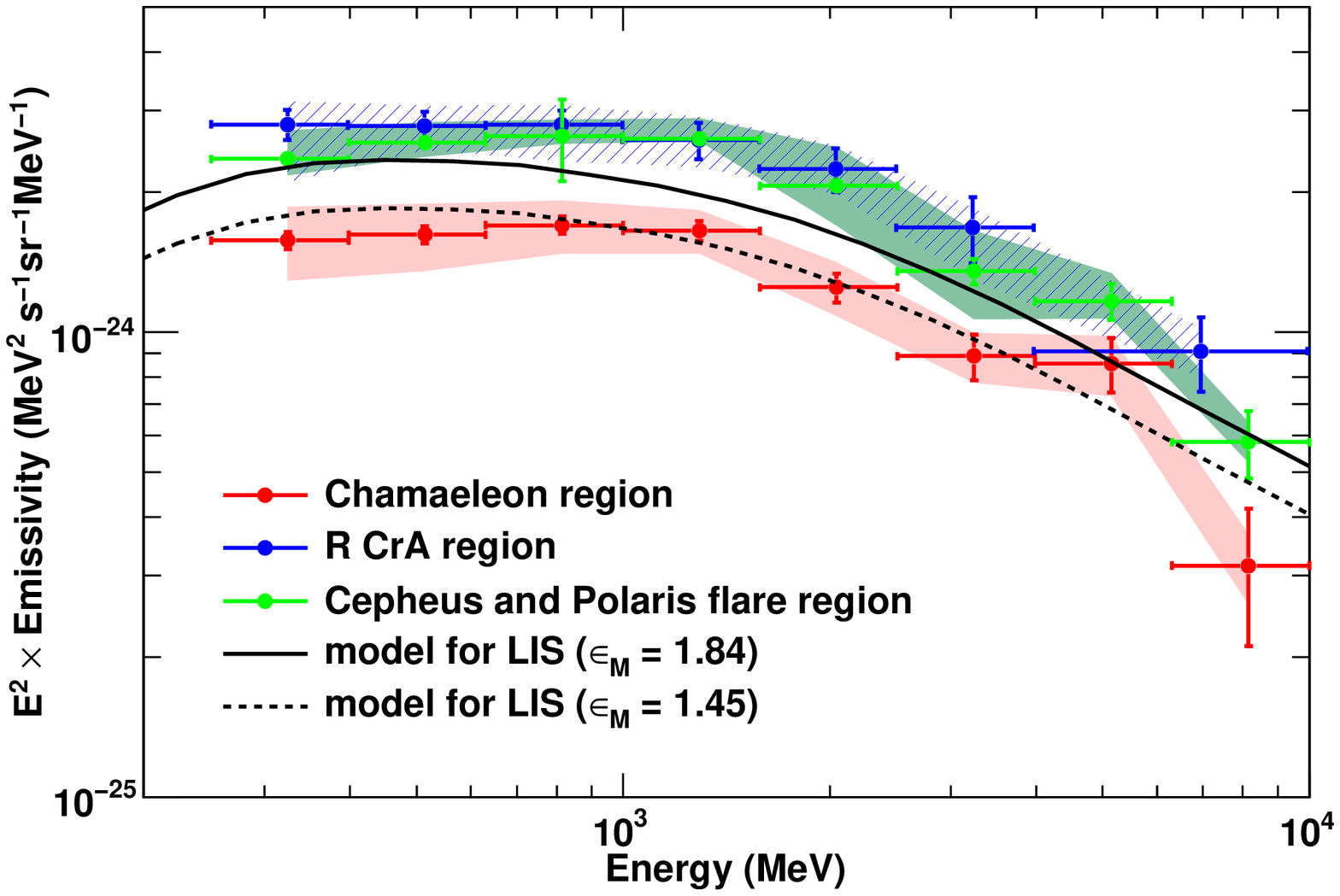}
  \end{center}
 \caption{{\HI} emissivity spectra of the Chamaeleon, 
 R CrA, and Cepheus and Polaris flare regions compared with the model for the LIS 
 with nuclear enhancement factors of 1.45 and 1.84. The shaded areas
 for the Chamaeleon, R CrA, and Cepheus and Polaris flare spectra indicate the 
 systematic uncertainty evaluated in Section \ref{sec:Results}.} 
 \label{fig:HI}
\end{figure}

It is generally believed that supernova remnants are the primary sources
of the Galactic CRs. Due to strong selection effects in detecting supernova remnants, 
their spatial distribution is not well determined. Therefore a smooth, axisymmetric
distribution has often been assumed in theoretical calculations,
resulting in a smooth decline of CR density as a function of the
Galactocentric radius. This assumption, however, should be tested
against observations. In Figure \ref{fig:HI}, we compare the obtained 
{\HI} emissivity spectra in the Chamaeleon, R CrA, and Cepheus and Polaris flare
regions (for {\it T}${\rm _S} =$ 125 K) and model emissivity 
spectra\footnote{The model is calculated from the LIS
compatible with the CR proton spectrum measured by Alcaraz et al. (2000)
and Sanuki et al. (2000), under the assumption that the nuclear
enhancement factors (the correction terms to take into account the contribution from 
nuclei heavier than protons in both CRs and interstellar matter) are 1.45 and 1.84 
(Mori 2009); see Abdo et al. (2009c).} 
for the local interstellar spectrum (LIS) used in Abdo et al. (2009c), 
based on local direct measurement of CRs. The spectral
shapes for the three regions studied here agree well with the LIS models,
indicating that the CR nuclei have similar spectral distribution in the
vicinity of the solar system. On the other hand, the absolute emissivities differ among regions. 
The emissivities of the three regions studied differ by $\sim$ 50\%,
estimated from the total $q_{\rm HI}$ shown in Tables 
\ref{table:fit_results_Cham}, \ref{table:fit_results_R_CrA}, and \ref{table:fit_results_CePo}.
We note that the systematic uncertainty of the LAT effective area
(5\% at 100 MeV and 20\% at 10 GeV; Rando et al. 2009) 
does not affect the relative value of emissivities
among these regions in Figure \ref{fig:HI}. Although the
emissivities of the R CrA region and the Cepheus and Polaris flare region are comparable,
that of the Chamaeleon region is lower by $\sim$ 20\%,
even if we take the systematic uncertainty into account (see Section \ref{sec:Results}).
As a further test, we fixed the emissivity of the Chamaeleon region
to that of the model for the LIS with the nuclear enhancement factor of 1.84
and performed the fitting. The fit turns out to be
significantly worse: the obtained ${\rm ln}(L)$ is lowered by 153 with
8 less free parameters. In addition, the normalization of the IC term is
lowered by more than a factor of three, although we cannot rule out 
such a low IC flux (low CR electron flux) for the direction of the Chamaeleon region.
The effect of unresolved point sources is small, since we
have verified that the obtained emissivities are robust against the
lower threshold for point sources between TS $=$ 50 and 100 (see
Section \ref{sec:Analysis_procedure}). 
We also confirmed that the residual excess of photons around 
($l = 280^{\circ}$ to $288^{\circ}$, $b = -20^{\circ}$ to $-12^{\circ}$; see the bottom panel of  
Figure \ref{fig:Cham_ana_map}) does not affect the local {\HI} emissivity very much.
Thus the total systematic uncertainty of the Chamaeleon region is 
conservatively estimated to be $\sim$ 15\% at most
(mainly due to the {\it T}${\rm _S}$, isotropic component and IC models), 
indicating a difference of the CR density between the Chamaeleon and
the others as shown in Figure \ref{fig:HI}. 

If the CR density has a variation by a factor of 1.2--1.5 in the 
neighborhood of the solar system, this requires a serious reconsideration 
of a smooth CR density often adopted for simplicity, and may have an impact 
on the study of the CR source distribution and diffuse $\gamma$-ray emission. 
We note that CR sources are stochastically distributed in space and time, 
and this may produce a CR anisotropy depending on
the propagation conditions as discussed by, e.g., Blasi \& Amato (2011a) and Blasi \& Amato (2011b). 
Study of other regions and more detailed theoretical calculations will be needed to further 
investigate this issue.

\subsection{Molecular Masses in the Interstellar Clouds Studied}

\begin{figure}
  \begin{minipage}{0.5\hsize}
   \includegraphics[width=80mm]{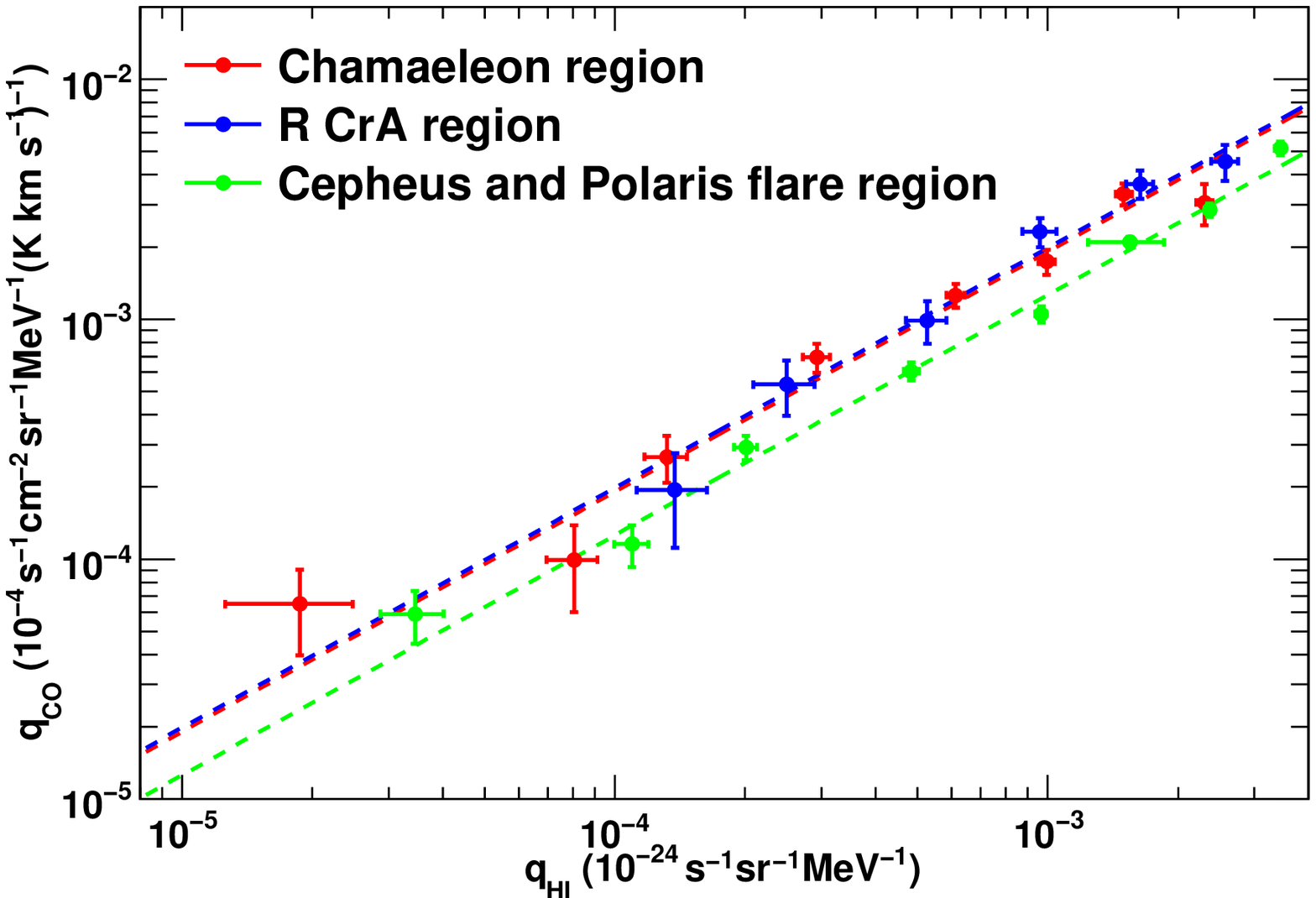}
   \end{minipage}
  \begin{minipage}{0.5\hsize}
   \includegraphics[width=80mm]{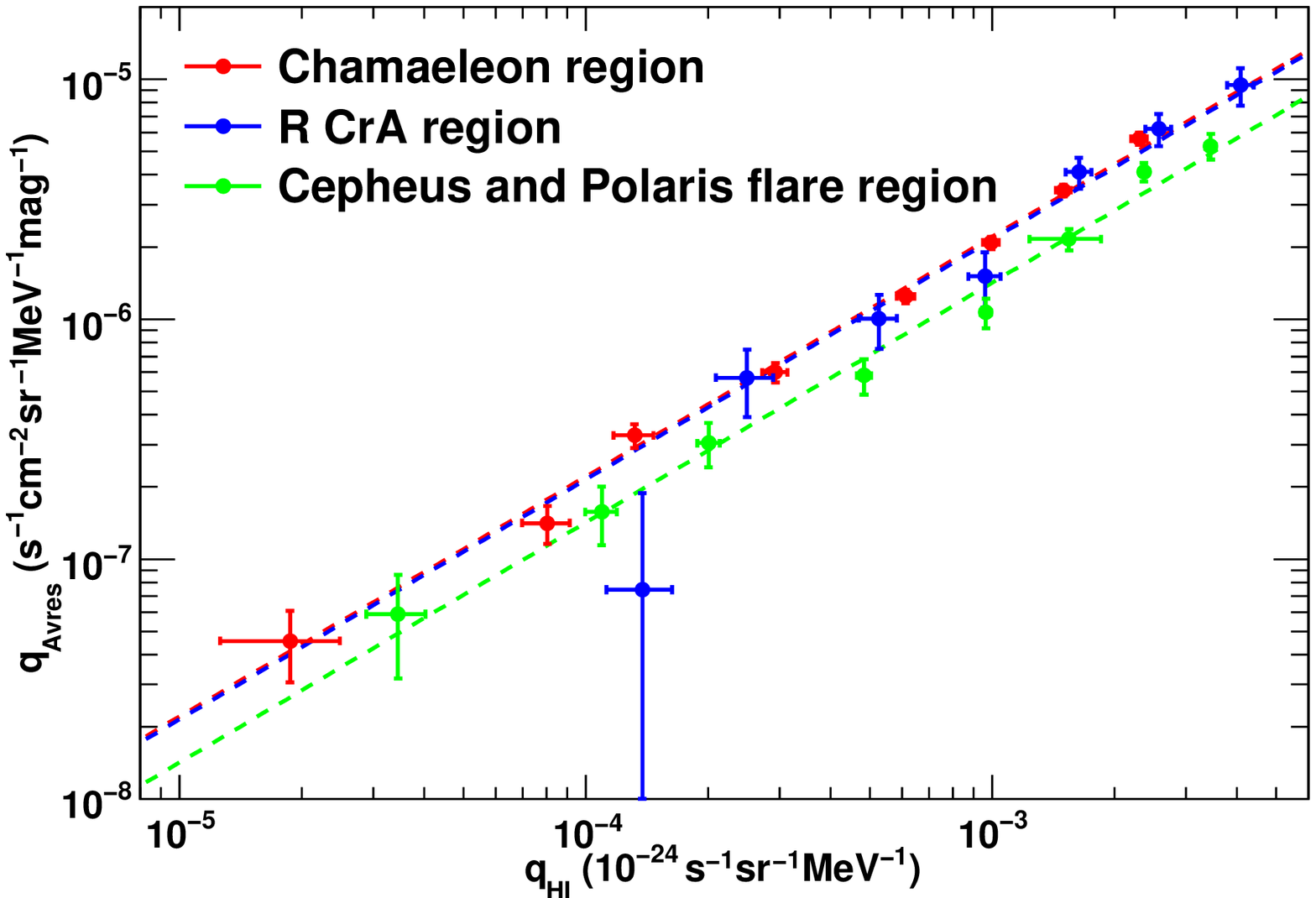}
  \end{minipage}
 \caption{CO (left) and $A{\rm v}_{\rm res}$ (right) versus {\HI} emissivities.
  Each point corresponds to an energy bin (see Tables \ref{table:fit_results_Cham}, 
  \ref{table:fit_results_R_CrA} and \ref{table:fit_results_CePo}).
  Errors are statistical only.} 
 \label{fig:Xco}
\end{figure}

Since the $\gamma$-ray production is almost independent of the chemical
or thermodynamical state of the interstellar gas, the $\gamma$-ray
observation is a powerful probe to investigate the molecular mass
calibration ratio, $X_{\rm CO}$, defined as $N({\rm H_{2}})/W_{\rm CO}$. 
Under the hypothesis that the same CR flux penetrates the {\HI} and CO phases of an
interstellar complex, we can calculate $X_{\rm CO}$ as $X_{\rm CO} = q_{\rm CO}/(2q_{\rm HI})$,
as shown in Figure \ref{fig:Xco} (left).
The linear relation supports the assumption that Galactic CRs penetrate
these molecular clouds uniformly to their cores. This also indicates
that any contamination from point sources and CR spectral variations in
molecular clouds analyzed here is small.

We have obtained $X_{\rm CO}$ by fitting the relation between 
$q_{\rm HI}$ and $q_{\rm CO}$ with a linear function using a
maximum-likelihood method; the $X_{\rm CO}$ values are
(0.96 $\pm$ 0.06$_{\rm stat}$ $^{+0.15}_{-0.12}$$_{\rm sys}$) $\times$ 10$^{20}$
cm$^{-2}$ (K km s$^{-1}$)$^{-1}$, 
(0.99 $\pm$ 0.08$_{\rm stat}$ $^{+0.18}_{-0.10}$$_{\rm sys}$) $\times$ 10$^{20}$
cm$^{-2}$ (K km s$^{-1}$)$^{-1}$, and
(0.63 $\pm$ 0.02$_{\rm stat}$ $^{+0.09}_{-0.07}$$_{\rm sys}$) $\times$ 10$^{20}$
cm$^{-2}$ (K km s$^{-1}$)$^{-1}$
for the Chamaeleon, R CrA, and Cepheus and Polaris flare regions, respectively.
The obtained value of $X_{\rm CO}$ for the Cepheus and Polaris flare region is
$\sim$ 20 \% lower than that reported by Abdo et al. (2010b).
Abdo et al. (2010b) includes in their study also the Cassiopeia molecular 
cloud in the Gould Belt, and due to different ROIs considered,
the $q_{\rm HI}$ emissivity was also different.
$X_{\rm CO}$ of the Chamaeleon region is similar to that of the R CrA region,
whereas that of the Cepheus and Polaris flare region is $\sim$ 2/3 of the 
others. The LAT data thus suggest a variation of $X_{\rm CO}$ on a $\sim$ 300 pc scale.

We can estimate the CO-bright molecular mass for the Chamaeleon, R
CrA, and Cepheus and Polaris flare regions. 
The mass of the gas traced by $W_{\rm CO}$ is expressed as
\begin{eqnarray}
\frac{M}{M_{\odot}} = 2 \mu\frac{m_{\rm H}}{M_{\odot}} d^{2} X_{\rm CO} \int
 W_{\rm CO}(l,b)\ d\Omega 
\label{eq:mass}
\end{eqnarray}
where $d$ is the distance to the cloud, $m_{\rm H}$ is the mass of the
hydrogen atom and $\mu = 1.36$ is the mean atomic weight per H-atom (Allen 1973). 
From this equation the mass of gas traced by CO
is expressed as $M_{\rm CO}$ in Table \ref{table:mass}:
we obtained $\sim$ 5$\times$10$^{3}$ $M_{\odot}$, $\sim$ 10$^{3}$ $M_{\odot}$, and
$\sim$ 3.3$\times$10$^{4}$ $M_{\odot}$
for the Chamaeleon, R CrA, and Cepheus and Polaris flare regions, respectively.   
The obtained mass of the Cepheus and Polaris flare region is $\sim$ 20 \% lower
than that reported by Abdo et al. (2010b) due to the different value of $X_{\rm CO}$.
Our estimates for the Chamaeleon and the R CrA regions 
are $\sim$ 1/2 of those by Dame et al. (1987);
they obtained $\sim$10$^{4}$ $M_{\odot}$ and $\sim$ 3$\times$10$^{3}$ $M_{\odot}$
and for the Chamaeleon and R CrA regions, respectively,
under the assumption of a high value of $X_{\rm CO}=$ 2.7 $\times$ 
10$^{20}$ cm$^{-2}$ (K km s$^{-1}$)$^{-1}$. 

Using the relation between $q_{\rm HI}$ and $q_{\rm Av}$, 
we can also calculate the mass of additional interstellar
gas traced by $A{\rm v}_{\rm res}$. Figure \ref{fig:Xco} 
shows the results of the fitting by a
linear relation, $q_{\rm Av} = X_{\rm Av} \cdot q_{\rm HI}$. 
The obtained $X_{\rm Av}$ values are 
(0.22 $\pm$ 0.01$_{\rm stat}$ $\pm$ 0.08$_{\rm sys}$) $\times$ 10$^{22}$
cm$^{-2}$ mag$^{-1}$, 
(0.21 $\pm$ 0.01$_{\rm stat}$ $\pm$ 0.02$_{\rm sys}$) $\times$ 10$^{22}$
cm$^{-2}$ mag$^{-1}$, and 
(0.14 $\pm$ 0.01$_{\rm stat}$ $\pm$ 0.03$_{\rm sys}$) $\times$ 10$^{22}$
cm$^{-2}$ mag$^{-1}$ 
for the Chamaeleon, the R CrA, and Cepheus and Polaris flare regions,
respectively. With the procedure similar to that for CO, we can calculate
the mass of additional gas traced by $A{\rm v}_{\rm res}$ 
expressed as $M_{\rm Av_{res}}$ in Table \ref{table:mass}:
we obtained $\sim$ 2.0$\times$10$^{4}$ $M_{\odot}$, $\sim$ 10$^{3}$ $M_{\odot}$ and
$\sim$ 1.3 $\times$ 10$^{4}$ $M_{\odot}$
for the Chamaeleon, R CrA, and Cepheus and Polaris flare regions, respectively.
We thus obtained mass estimates for the Chamaeleon and R CrA regions 
similar to previous ones (Dame et al. 1987)
if we consider the total mass (traced by $W_{\rm CO}$ and $A{\rm v}_{\rm res}$),
although the procedure is not straightforward since the gas traced by
$A{\rm v}_{\rm res}$ is extended in a much larger region of the
sky. Detailed study of the matter distribution in the interstellar space
by comparing $\gamma$-rays and other tracers will be reported elsewhere.

\begin{table}[t]
 \caption{\normalsize{Masses in the interstellar clouds for each region.}}
 \label{table:mass}
  \begin{center}
   \begin{tabular}{cccccc}\hline\hline
   Region     & $l$         & $b$       & $d$ (pc) & $M_{\rm CO}$ ($M_{\odot}$)
   &  $M_{\rm Av_{res}}$ ($M_{\odot}$)\\ \hline
   Chamaeleon & [295$^{\circ}$, 305$^{\circ}$]  & [--20$^{\circ}$, --12$^{\circ}$] & 215$^{a}$ 
   & $\sim$ 5$\times$10$^{3}$ &
   $\sim$ 2.0$\times$10$^{4}$ \\
   R CrA      & [--1$^{\circ}$, 4$^{\circ}$]     & [--24$^{\circ}$, --16$^{\circ}$] & 150$^{a}$ 
   & $\sim$ 10$^{3}$ & 
   $\sim$ 10$^{3}$ \\
   Cepheus and Polaris flare & [100$^{\circ}$, 125$^{\circ}$] & [15$^{\circ}$, 30$^{\circ}$] &
   300$^{b}$ 
   & $\sim$ 3.3 $\times$ 10$^{4}$ & 
   $\sim$ 1.3 $\times$ 10$^{4}$ \\ \hline
   \multicolumn{6}{l}{{\bf Notes.}$^{a}$ Dame et al. (1987), $^{b}$ Abdo et al. (2010b)}\\
   \end{tabular}
  \end{center}
\end{table}

\clearpage

\section{Summary and Conclusions}
\label{sec:Summary_and_conclusions}

We have studied the $\gamma$-ray emission from the Chamaeleon, R CrA,
and Cepheus and Polaris flare molecular clouds close to the solar system 
($\lesssim$ 300 pc) using the first 21 months of {\it Fermi} LAT data. 
Thanks to the excellent performance of the LAT, we have obtained 
unprecedentedly high-quality emissivity spectra of the atomic and 
molecular gas in these regions in the 250 MeV -- 10 GeV range.

The $\gamma$-ray emissivity spectral shapes in three regions 
agree well with the model for the LIS (a model based on local CR measurement),
thus indicating a similar spectral distribution of CRs in these regions. 
The emissivities, however, indicate a variation of the CR 
density of $\sim$ 20 \% within $\sim$ 300 pc around the solar system,
even if we consider the systematic uncertainties.
We consider possible origins of the variation are non-uniform
supernova rate and anisotropy of CRs depending on the propagation conditions.  

The molecular mass calibration ratio $X_{\rm CO}$ for 
the Chamaeleon cloud and the R CrA cloud are comparable, 
whereas that of the Cepheus and Polaris flare region is $\sim$ 2/3 of the others, 
suggesting a variation of $X_{\rm CO}$ in the vicinity of the solar system. 
From the obtained values of $X_{\rm CO}$, the masses of gas traced by 
$W_{\rm CO}$ in the Chamaeleon, R CrA, and Cepheus and Polaris flare regions 
are estimated to be $\sim$ 5 $\times$ $10^{3}$ $M_{\odot}$, 
$\sim10^{3}$ $M_{\odot}$, and $\sim$ 3.3 $\times$ $10^{4}$ $M_{\odot}$
respectively. Similar amounts of gas are inferred to be in the 
phase not well traced by the {\HI} or CO lines. Accumulation of more 
$\gamma$-ray data, particularly at high energies, and progress in ISM 
studies, will reveal the CR and matter distribution in greater detail. 

The {\it Fermi} LAT Collaboration acknowledges generous ongoing support
from a number of agencies and institutes that have supported both the
development and the operation of the LAT as well as scientific data
analysis. These include the National Aeronautics and Space
Administration and the Department of Energy in the United States, the
Commissariat $\grave{\rm a}$ l'Energie Atomique and the Centre National de
la Recherche Scientifique/Institut National de Physique Nucl$\acute{\rm e}$aire
et de Physique des Particules in France, the Agenzia Spaziale Italiana
and the Istituto Nazionale di Fisica Nucleare in Italy, the Ministry of
Education, Culture, Sports, Science and Technology (MEXT), High Energy
Accelerator Research Organization (KEK), and Japan Aerospace Exploration
Agency (JAXA) in Japan, and the K. A. Wallenberg Foundation, the Swedish
Research Council, and the Swedish National Space Board in Sweden.

Additional support for science analysis during the operations phase is
gratefully acknowledged from the Istituto Nazionale di Astrofisica in
Italy and the Centre National d'$\acute{\rm E}$tudes Spatiales in France.

We thank the GALPROP team for providing a development version of GALPROP
model adjusted to the measurement data by the LAT. GALPROP development is
supported by NASA Grant NNX09AC15G and by the Max Planck Society.

\end{document}